\documentclass[prb,showpacs,preprintnumbers,amsmath,amssymb,twocolumn,superscriptaddress,longbibliography]{revtex4-1}

\usepackage[dvipdfmx]{graphicx}
\usepackage{dcolumn}
\usepackage{bm}
\usepackage{color}

\usepackage{amsmath}	
\usepackage{amsfonts}
\usepackage{braket}
\usepackage{amsthm}

\usepackage{comment}

\newtheorem{claim}{Claim}

\begin{document}

\title{
Projectification of point group symmetries with a background flux and
Lieb-Schultz-Mattis theorem
}

\author{Yasuhiro Tada}
\email[]{ytada@hiroshima-u.ac.jp}
\affiliation{
Quantum Matter Program, Graduate School of Advanced Science and Engineering, Hiroshima University,
Higashihiroshima, Hiroshima 739-8530, Japan}
\affiliation{Institute for Solid State Physics, University of Tokyo, Kashiwa 277-8581, Japan}

\author{Masaki Oshikawa}
\affiliation{Institute for Solid State Physics, University of Tokyo, Kashiwa 277-8581, Japan}

\begin{abstract}
We discuss the Lieb-Schultz-Mattis (LSM) theorem in two-dimensional spin systems with on-site 
${\mathrm U}(1)\rtimes {\mathbb Z}_2$ spin rotation symmetry and point group $C_{2v}$
symmetry about a site.
We ``twist" the point group symmetry by introducing a small uniform U(1) flux
to obtain a projective symmetry, similarly to the familiar magnetic translation symmetry.
The LSM theorem is proved in presence of the flux and then it is demonstrated that the theorem
holds also for the flux-free system.
Besides, the uniform flux enables us to show the LSM theorem for the time-reversal symmetry
and the site-centered $C_2$-rotation symmetry.
\end{abstract}

\maketitle

\section{introduction}
Symmetry of a system is a fundamental property and often responsible for structures of 
a low energy spectrum.
One of the most celebrated examples is the Nambu-Goldstone theorem, and 
there exist gapless excitations if a continuous symmetry is spontaneously broken
~\cite{Nambu1960,Nambu1961,Goldstone1961,Goldstone1962}.
Existence or absence of
such low energy excitations is crucial for
basic behaviors of physical quantites, and usually 
gapless excitations lead to power law behaviors of an observable while
exponential behaviors are seen for gapped excitations~\cite{HastingsKoma2006}.
Furthermore, a low energy spectrum with an excitation gap above the ground state can characterize
different classes of quantum phases.
The ground state $\ket{\Psi}$ is a symmetry protected topological state when $\ket{\Psi}$
is unique and gapped,
while $\ket{\Psi}$ is a topologically ordered state or a state with a spontaneously broken symmetry
when the ground states are degenerate.
In general, however, it is difficult to obtain energy eigenvalues of a given Hamiltonian,
and calculations of a low energy spectrum is a central problem in condensed matter physics.

In this context, the Lieb-Schultz-Mattis (LSM) theorem is a fundamental no-go theorem 
which imposes a strong constraint on possible energy spectra based on symmetries
~\cite{LSM1961,AffleckLieb1986,Oshikawa2000,Hastings2004,NS2007,Bachmann2020,Cho2017,
Parameswaran2013,Watanabe2015,Yao2020,Furuya2019,Yao2021,CGW2011,Fuji2016,OTT2020,Tasaki2022book,
Koma2000,Lu2020,Lu2024,Tada2021,Yao2022,Po2017,Else2020,Huang2017,Qi2017,Yao2024}.
One can exclude the possibility of a unique ground state with an excitation gap based only on symmetries 
of the system 
without explicit diagonalization of  the Hamiltonian.
The LSM theorem was originally proved for a one-dimensional
spin system with the on-site U(1) spin rotation symmetry and
translation symmetry~\cite{LSM1961,AffleckLieb1986}.
It is important to extend the LSM theorem to systems with spatial symmetries other than the translation symmetry.
To this end, let us briefly review the existing approaches to the LSM theorem.

The original proof of the LSM theorem in one dimension is based on a variational argument~\cite{LSM1961,AffleckLieb1986};
the LSM ``twist'' operator is applied to the ground state to create a low energy excited state.
Then the orthogonality of the two state is derived from the nontrivial commutation relation between the twist operator
and the lattice translation operator.
Furthermore, the twist operator is nothing but the large gauge transformation operator.
Combined with an adiabatic flux insertion argument, the commutation relation between the twist (large gauge transformation) operator
and the lattice translation operator can be used to show the LSM theorem in higher dimensions~\cite{Oshikawa2000}.

There is also an alternative approach to the LSM theorem: first consider the system with a ``twisted'' boundary condition~\cite{Yao2021}.
In the presence of a twisted link, the Hamiltonian is no longer invariant under the lattice translation.
Nevertheless, physically the system is uniform since the location of the twist is arbitrary.
As a consequence, the Hamiltonian is invariant under a combination of the lattice translation and a local gauge transformation.
The dressed translation operator may have a nontrivial commutation relation with a global on-site symmetry operation,
leading to the exact ground-state degeneracy.
In this way, under certain conditions, we can show exact ground-state degeneracy under the twisted boundary condition.
Since the bulk property of the system should be independent of the boundary condition, we can expect quasi-degeneracy
of ground states under the periodic boundary condition for a very large system (which is the usual statement of the LSM theorem).
This approach has been used to extend the LSM theorem to systems with discrete internal symmetries where the slow twist
or the adiabatic flux insertion argument is not applicable.

The boundary twisting can be also applied to  higher dimensional systems with the translation symmetry.
However, this approach is not generally useful to derive the LSM theorem for point-group symmetries.
The point group symmetry leads to exact degeneracy of the ground states under a twisted boundary condition~\cite{Yao2022,Yao2024},
but it is subtle to derive quasi-degeneracy under the periodic boundary condition when there is no 
additional translation symmetry.
In contrast to a translation symmetric system, the location of the twist is not arbitrary in a 
non-uniform system only with the point group symmetry 
and untwisting the twisted boundary on the chosen link might possibly affect the energy spectrum.

The point group LSM theorem has been discussed based on other various approaches without any boundary twist~\cite{Parameswaran2013,Watanabe2015,CGW2011,Fuji2016,OTT2020,Po2017,Else2020,Huang2017,Qi2017}. 
A basic idea behind is to introduce a quantity or an index which forbids a uniquely gapped ground state,
In one dimension, the LSM theorem for the inversion symmetry has been established even in a mathematically 
rigorous manner~\cite{OTT2020}.
For two-dimensional systems, several arguments have been proposed to show the LSM theorem
in the presence of the point group symmetry such as $D_2$~\cite{Huang2017,Qi2017}.
However, each of them relies on a nontrivial assumption (such as extensive trivialization by a symmetric local unitary).
In order to deepen our understanding of the LSM theorem for point group symmetries
and for further expansions, it is desirable to develop an alternative approach.

In this work, we propose a framework to ``projectify'' a point group symmetry in two-dimensional systems
by introducing a small uniform U(1) flux perpendicular to the two-dimensional plane.
The introduction of the flux plays a similar role to those of the boundary twist and 
leads to a projective representation of the point group similarly to
the familiar magnetic translation group~\cite{Koma2000,Lu2020,Lu2024,Tada2021},
naturally leading to an exact ground-state degeneracy.
Since the background flux density approaches to zero in the thermodynamic limit,
we can expect that the spectrum of the flux-free Hamiltonian is almost unchanged;
thus a quasi-degeneracy of the ground states should remain in the flux-free system,
amounting to the LSM theorem.
Our framework provides a new perspective and physical insight into the LSM theorem
for point-group symmetries.

This paper is organized as follows.
We first discuss one-dimensional systems as a starting point in  Sec.~\ref{sec:uniform}
and examine two-dimensional systems based on the one-dimensional case in  Sec.~\ref{sec:twodim}. 
For two-dimensional systems, 
we introduce a tiny uniform U(1) flux perpendicular to the plane to obtain a projective representation
of the point group, which leads to exact degeneracy of eigenvalues of the Hamiltonian under the flux.
It is shown that the energy spectrum is almost unchanged by the tiny flux,
and therefore the LSM theorem holds for the flux-free Hamiltonian.
Finally, a summary is given in Sec.~\ref{sec:summary}.
For a comparison,
the LSM theorem from the locally twisted point group symmetry is also discussed
in Appendix.

\section{One dimensional system}
\label{sec:uniform}
The main target of this study is two dimensional systems with
the point group $C_{2v}$ (or equivalently $D_2$ in two dimensions).
We prove the LSM theorem for half-integer spin systems
with the on-site ${\mathrm U}(1)\rtimes{\mathbb Z}_2$ and point group symmetries
by introducing a tiny uniform U(1) flux.
The same approach also enables us to show the LSM theorem with use of 
the time-reversal symmetry.
We prove several statements based on symmetries and call them LSM ``theorems" 
for simplicity, although they are not mathematically rigorous.

Before discussing two dimensional systems, we briefly discuss the one-dimensional LSM theorem
in a half-integer spin chain with the U(1)$\rtimes {\mathbb Z}_2$ spin rotation symmetry and the ${\mathbb Z}_2^T$ time-reversal symmetry 
in presence of 
the site-centered inversion symmetry~\cite{AffleckLieb1986,Yao2024},
in our framework of introducing a background flux.
Although in this section we will only re-derive a subset of the previously known LSM theorems in one dimension,
it will serve as a warm-up for the two-dimensional case.
In one dimensional chain, we can only introduce the Aharonov-Bohm flux for a periodic boundary condition
as a nontrivial flux.
This is rather different from the weak background flux piercing each plaquette in two dimensions,
which we will utilize later.
Nevertheless, there is a parallel between the two cases as we will see in the following.

We consider, as a prototypical example, the half-integer spin XXZ model defined on a lattice $x_j=0,1,2,\cdots, L-1$
with the periodic boundary condition,
\begin{align}
H=\sum_{i}\frac{J_i}{2} (S^+_iS^-_{i+1}+S^-_iS^+_{i+1})+J_i^zS^z_iS^z_{i+1},
\label{eq:H1D}
\end{align}
where the magnetic coupling $J_i$ is real.
The on-site U(1)$\rtimes{\mathbb Z}_2$ spin rotation symmetry is described by the
unitary operators,
\begin{align}
R^x_{\pi}=e^{i\pi \sum_jS_j^x}, \quad R^z_{\theta}=e^{i\theta\sum_jS_j^z},
\label{eq:R}
\end{align}
where $\theta$ is a rotation angle.
Besides, the system has the anti-unitary time-reversal symmetry described by the operator
\begin{align}
T=R^y_{\pi}K, \quad R^y_{\pi}=e^{i\pi\sum_jS_j^y},
\label{eq:TR}
\end{align}
where $R^y_{\pi}$ is the $\pi$-rotation of spins about the $y$-axis
and $K$ is the complex conjugation operator.
Although the XXZ model has both of spin-rotation and time-reversal symmetries, 
more general models can have only one symmetry if
there exists a term which breaks one of them such as $i(S_j^+S_k^--S_j^-S_k^+)$.
We will discuss LSM theorems corresponding to one of these two symmetries separately.

The internal symmetry has the commutation relation, $R^x_{\pi}R^z_{\pi}
=(-1)^{L}R^z_{\pi}R^x_{\pi}$ with
the system size $L$, and it gives a non-trivial commutation relation when $L$ is odd
similarly to the time-reversal symmetry.
To avoid such trivial degeneracy and obtain non-trivial one based on the spatial symmetry, 
the system size $L$ is assumed to be even.
(Note that, in general, there are systems where the total number of spins must be even such as the checkerboard lattice;
this will be discussed later in Sec.~\ref{sec:extension}.)
In this case, the $\pi$-rotations are squared to unity, $(R^x_{\pi})^2=(R^z_{\pi})^2=1$.
For spatial symmetry,
we suppose that 
$J_i,J_i^z\in{\mathbb R}$
are not necessarily one site translationally invariant, 
but the system has the inversion symmetry $I$ about the origin $x_0=0$,
\begin{align}
I S_{j}^{\mu} I^{-1}=S_{L-j}^{\mu}, \quad \mu=x,y,z.
\end{align}
Then combinations of the on-site symmetry and the inversion symmetry lead to the LSM theorems.
The on-site symmetry can be either the unitary ${\mathbb Z}_2$ spin rotation or
the anti-unitary ${\mathbb Z}_2^T$ time-reversal symmetry in addition to the U(1) spin rotation symmetry.
The statements are as follows.
\begin{claim}  [Affleck-Lieb~\cite{AffleckLieb1986}]
A one-dimensional
half-integer spin XXZ model with the on-site ${\mathrm U}(1)\rtimes{\mathbb Z}_2$ and 
site-centered inversion symmetries
does not have a unique gapped ground state under the periodic boundary condition.
\label{thm1}
\end{claim}
\begin{claim}
A one-dimensional
half-integer spin XXZ model with the on-site ${\mathrm U}(1)\times{\mathbb Z}_2^T$ and 
site-centered inversion symmetries
does not have a unique gapped ground state under the periodic boundary condition.
\label{thm1TR}
\end{claim}
Claim~\ref{thm1} is a simplified version of the rigorous theorem proved by Affleck and Lieb~\cite{AffleckLieb1986}
and we provide a proof in Appendix~\ref{app:proof1}.
Claim~\ref{thm1TR} is a weak variant of the LSM theorem discussed in the previous study~\cite{Yao2024}. 
Note that 
the bond-centered inversion symmetry does not lead to an LSM theorem.
A simple example is the Hamiltonian $H=J\sum_{k=0}^{L/2-1}\vec{S}_{2k}\cdot \vec{S}_{2k+1}$ with $J>0$
which has the unique gapped ground state 
$\ket{\Psi}=\otimes\ket{\mbox{bond singlet}}$.
On the other hand, the site-centered inversion symmetry plays an essential role as shown in the following.

{\it Proof of Claim~\ref{thm1TR}}.
For the present XXZ model,
we first insert a U(1) flux $\Phi_x=-\pi$ through the non-contractible loop of the system,
and consider the Hamiltonian $H(\Phi_x)$ under the flux.
To be precise, we introduce a uniform gauge field $A_x=\pi/L$ and modify the coupling constant 
as $J_{j}\to J_{j}e^{iA_x}$.  
Then, the Hamiltonian $H(\Phi_x)$ has the time-reversal symmetry and also a modified inversion symmetry.
The twisted inversion operator is given by
\begin{align}
\tilde{I}&=IU_{x},\\
U_{x}&=\exp\left(i\frac{2\pi}{L}\sum_{j}x_jS_j^z\right),
\label{eq:U1D}
\end{align}
The operator $U_x$ is often called Lieb-Schultz-Mattis twsting operator~\cite{LSM1961,AffleckLieb1986}
corresponding to the U(1) symmetry.
It is crucial that $\tilde{I}$ satisfies $\tilde{I}^2=-1$, because $\tilde{I}^2=\exp(i2\pi\sum_{j\neq0}S_j^z)=e^{i\pi}$
for an even $L$.
Eigenvalues of ${\tilde I}$ are imarginary $\pm i$, which is in contrast to the original $I$-operator whose
eigenvalues are real $\pm1$.
Clearly, the two symmetry operators commute each other,
\begin{align}
\tilde{I} T=T\tilde{I}.
\label{eq:IT}
\end{align}
This means that, given a simultaneous eigenstate $\ket{\Psi}$ of $H(\Phi_x)$ and $\tilde{I}$,
the partner state $\ket{\Phi}=T\ket{\Psi}$ has a different eigenvalue of $\tilde{I}$.
(The state $\ket{\Phi}$ is degenerate with $\ket{\Psi}$ because of the time-reversal symmetry
of the twisted Hamiltonian, $[H(\Phi_x),T]=0$.)
Thus, the two states are orthogonal, $\langle \Psi|\Phi\rangle=0$,
which implies there is a pair of exactly degenerate states for the Hamiltonian $H(\Phi_x)$.
Next, we want to extrapolate this result to a flux-free system $H(0)$. 
The robustness of a spectral gap against a U(1) flux has not been fully proved~\cite{Oshikawa2000,Watanabe2018},
but it is widely considered that U(1) symmetry twisting does not open an energy gap.
Here, we simply assume that this is true in the present one-dimensional system, namely,
ground states of $H(0)$ are not uniquely gapped if those of $H(\Phi_x)$ are not uniquely gapped.
Then, the LSM theorem for the Hamiltonian $H(0)$ immediately follows from the exact ground state degeneracy for $H(\Phi_x)$.

The above discussion consists of two steps, where we showed that 
(i) ground states are degenerate for a twisted Hamiltonian and 
(ii) argued that the degeneracy remains even for the untwisted original Hamiltonian.
Similar two-step approaches have been used for systems
with translation symmetry~\cite{Tada2021} and also for multipole insulators with point group symmetry~\cite{Tada2023}. 
Although the step (ii) is an assumption in the above one dimensional case,
it truns out that
the extrapolation can be done in a controlled way in two dimensional systems.
The two-step argument will be repeatedly applied in the next section.

\section{Two dimensional system}
\label{sec:twodim}
We extend the previous argument for a one-dimensional system
to a system defined on a two-dimensional lattice $\Lambda$ with the size $L_x\times L_y$.
An important point in the proof of Claim~\ref{thm1TR} is that 
the inversion symmetry is twisted by the U(1) flux along the cylinderical direction of the system.
In this study, we introduce a small uniform U(1) flux $\phi$ perpendicular to the two-dimensional plane 
to construct a suitable operator 
and controll the low energy states in an efficient manner,
where the on-site and spatial symmetries are correlated by the flux.
(For a comparison, we also briefly discuss symmetry operators under a local flux in Appendix~\ref{sec:local}.)
The point group symmetry of the system gets
twisted by a phase factor and it becomes projective, similarly to the familiar magnetic translation 
symmetry\cite{Koma2000,Lu2020,Lu2024,Tada2021}. The projective symmetry leads to a pair of the degenerate ground states for the Hamiltonian with the flux.
Then, we extrapolate the flux $\phi\to0$ to discuss the original flux-free system. 
(``Flux-free" simply means that there is no flux in the Hamiltonian.)
In this way, we prove the following statements.
\begin{claim}
A two-dimensional 
half-integer spin XXZ model with the on-site ${\mathrm U}(1)\rtimes{\mathbb Z}_2$ and point group $C_{2v}$ symmetries 
does not have a unique gapped ground state under the periodic boundary condition.
\label{thm2}
\end{claim}
\begin{claim}
A two-dimensional 
half-integer spin XXZ model with the on-site ${\mathrm U}(1)\times{\mathbb Z}_2^T$ and $C_2$-rotation symmetries 
does not have a unique gapped ground state under the periodic boundary condition.
\label{thm2TR}
\end{claim}
The $C_{2v}$ point group symmetry (or equivalently $D_2$ in two dimensions) is required in the former, while $C_2$-rotation alone
is sufficient in the latter.
The point group is supposed to have a fixed site and also the $C_2$-rotation is site-centered.
Note that bond- or plaquette-centered point group symmetry does not lead to an LSM theorem
similarly to the one-dimensional systems with the bond-centered inversion symmetry. 
The site-centered point group symmetry plays essential roles as will be discussed in the following sections.
A similar statement to Claim~\ref{thm2} was discussed previously, 
where the action of the point group symmetry on a spin itself is projective in the absence of an additional flux
~\cite{Huang2017,Qi2017}.
In contrast, the action of the point group symmetry itself (Eq.~\eqref{eq:gaction}) is linear in the present study, 
and it is twisted by the U(1) flux to give a projective representation.

To show the above statements, we first consider a simple model in Sec.~\ref{sec:simple1}.
We consider the XXZ model on a square lattice and introduce a small U(1) flux to twist the point group symmetry
by the on-site symmetry.
Then, we explicitly construct the projective point group symmetry under the flux.
We prove Claim~\ref{thm2} based on the projective symmetry.
Another but equivalent set of projective operators in a different gauge field is given in Sec.~\ref{sec:simple2}.
Claim~\ref{thm2TR} will be discussed in Sec.~\ref{sec:uniformTRS}.
Finally, we extend our argument to general cases in Sec.~\ref{sec:extension} to complete the proof.

\subsection{XXZ model on square lattice}
\label{sec:simple1}
{\it Model.}
We consider the half-integer spin XXZ model on the square lattice as a simple protoypical example,
\begin{align}
H=\sum_{ij}\frac{J_{ij}}{2} (S^+_iS^-_j+S^-_iS^+_j)+J_{ij}^zS^z_iS^z_j,
\label{eq:H2D0}
\end{align}
where the magnetic coupling is real and $J_{ij}\neq0$ only for the nearest neighbors.
The periodic boundary condition has been imposed for both directions.
We will discuss some extensions later in Sec.~\ref{sec:extension}.
The model has the on-site U(1)$\rtimes{\mathbb Z}_2$ symmetry 
described by the operators $R^x_{\pi}$ and $R^z_{\theta}$ similarly to the one-dimensional model.
In addition, we assume that the system has the point group symmetry of the square lattice,
$C_{4v}=\{1,C_4,C_{4}^{-1},C_2,M_{x},M_{y},M_{xy},M_{\bar{xy}}\}$, 
where $M_{x(y)}$ is the mirror about the $x(y)$-axis and $M_{xy(\bar{xy})}$ 
is the mirror about the $x=y$ $(x=-y)$ line.
To avoid the trivial degeneracies arising from the spin rotation symmetry with 
$R^x_{\pi}R^z_{\pi}=(-1)^{|\Lambda|}R^z_{\pi}R^x_{\pi}$
and the time-reversal symmetry $T^2=(-1)^{|\Lambda|}$, 
the lattice size is taken to be $L_x=L_y\equiv L\in 2{\mathbb Z}$. 
In this geometry, there are mirror symmetries $M_{xy}, M_{\bar{xy}}$ between the $x,y$-axes
in addition to the mirror symmetries $M_{x},M_y$ about each axis.

\begin{figure}[tbh]
\includegraphics[width=5.0cm]{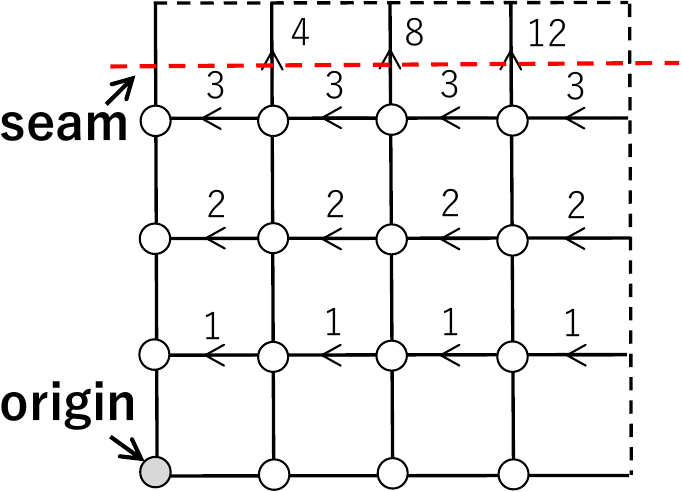}
\caption{The Landau string gauge for a $L_x=L_y=L=4$ square lattice where circles represent lattice sites and
the periodic boundary condition has been imposed.
Each number on the bonds corresponds to $A_{ij}$ in unit of $\phi_L=2\pi /L^2$
and its $y$-component is non-zero only when a bond crosses the ``seam" on the top plaquettes
represented by the red dashed line.
The site at the left-bottom corner is defined as the origin $(x,y)=(0,0)$.
The Landau string gauge can be defined for arbitrary system sizes.
}
\label{fig:string1}
\end{figure}

The action of the point group is given by
\begin{align}
gS^{\mu}_jg^{-1}=S^{\mu}_{g^{-1}j}, \quad g\in C_{4v}.
\label{eq:gaction}
\end{align}
Note that, in contrast to the previous studies~\cite{Huang2017,Qi2017},
the spin and the space are not directly correlated and the action of $C_{4v}$ is linear in the present study 
where spin-orbit interactions are not taken into account.
We can also consider a projective representation of the point group for each spin, 
but it is reduced to a linear representation
of the group action for the total system with the even number of the half-integer spins.
Therefore, we simply use the above definition Eq.~\eqref{eq:gaction} in this study, for which 
it is clear that projectiveness comes only from the additional flux.
For our discussion on the LSM theorem,
we focus on the Abelian subgroup of $C_{4v}$ consisting of the $C_2$-rotation and two mirrors,
$C_{2v}=\{1,C_2,M_1,M_2\}$,  where $\{M_1,M_2\}$ is either $\{M_x,M_y\}$ or $\{M_{xy}, M_{\bar{xy}}\}$.
To twist the trivial commutation relation $g_1g_2=g_2g_1$ of $g_1,g_2\in C_{2v}$
by the U(1) symmetry, we introduce the uniform U(1) flux for each plaquette.
  The total flux perpendicular to the two-dimensional system is quantized in units of $2\pi$,
  under the periodic boundary conditions.
  Thus, the smallest non-zero uniform flux per plaquette in a system of linear size $L$ is given by
\begin{equation}
\phi_L=\frac{2\pi}{L^2}.
\label{eq:phi_L}
\end{equation}
To realize this minimal uniform flux, we introduce
the corresponding gauge field $A_{jk}$ so that $\sum_{jk\in p}A_{jk}=\phi_L$ for a plaquette $p$.
The Hamiltonian under the flux reads
\begin{align}
H(\phi_L)=\sum_{jk}
\frac{J_{jk}}{2}(e^{iA_{jk}}S^+_jS^-_k+e^{-iA_{jk}}S^-_jS^+_k)+J_{jk}^zS^z_jS^z_k.
\label{eq:H2DA}
\end{align}
The vector potential $A_{jk}$ is given in the string gauge 
and we fix the concrete configuration as shown in Fig.~\ref{fig:string1}~\cite{Tada2021,Hatsugai1999,Kudo2017}.
This can be considered as a variant of the Landau gauge on the lattice torus,
where the Landau gauge is $\vec{A}=B(-y,0,0)$ in the Euclidian space.
We call it the Landau string gauge.
Note that the string gauge $A_{jk}$ looks a bit complicated, but it realizes the smallest uniform flux 
$\phi_L=2\pi/L^2$ for every plaquette $p$ and total flux $\sum_{p}\phi_p=2\pi$ for the entire system
which is consistent with the periodic boundary condition.
Besides, the Landau string gauge respects $M_{x(y)}$-mirror among the four mirror symmetires of $C_{4v}$
as will be shown later.
We also note that the twisted Hamiltonian Eq.~\eqref{eq:H2DA} preserves the time-reversal symmetry $T$.
This is because the present U(1) flux is a spin gauge field which preserves the time-reversal
symmetry, $T(e^{iA_{jk}}S_j^+S_k^-+e^{-iA_{jk}}S_j^-S_k^+)T^{-1}=e^{-iA_{jk}}S_j^-S_k^++e^{iA_{jk}}S_j^+S_k^-$.
On the other hand, the $\pi$-rotations $R^x_{\pi}, R^y_{\pi}$ themselves flip the spin flux
$+\phi\to-\phi$, since they are unitary and $R^x_{\pi}S_j^+(R^x_{\pi})^{-1}=S_j^-$.

{\it Construction of projective point group symmetry.}
To obtain a projective symmetry of $C_{2v}$,
we first consider $C_2$-rotation about the origin, $C_2:(x_j,y_j)\to(L-x_j,L-y_j)$,
and suppose that it is written in the form of 
\begin{align}
\tilde{C}_2=C_2U_2, \quad U_2=e^{i\sum_j\xi_j S_j^z},
\end{align}
where  $\xi_j$ is a scalar function to be determined.
Then the $S^+S^-$ and $S^-S^+$ terms in the Hamiltonian are changed by this operator as
\begin{align}
\tilde{C}_2e^{iA_{jk}}S^+_jS^-_k\tilde{C}_2^{-1}= e^{iA_{jk}+i\xi_j-i\xi_k}S^+_{C_2j}S^-_{C_2k}.
\label{eq:SpSm}
\end{align}
To realize the symmetry $[\tilde{C}_2,H(\phi_L)]=0$, the transformed term Eq.~\eqref{eq:SpSm} must be identical to 
$\exp(iA_{C_2j,C_2k})S^+_{C_2j}S^-_{C_2k}$, namely
\begin{align}
A_{C_2j,C_2k}=A_{jk}+d\xi_{jk},\quad d\xi_{jk}=\xi_j-\xi_k.
\end{align}
This is a difference equation of $\xi_j$ and
one can easily obtain a solution, 
\begin{align}
\xi_j=\frac{2\pi}{L}x_j(1-\delta_{y_j,0})
\end{align}
for $0\leq x_j,y_j\leq L-1$.
The gauge transformation operator $U_2$ with the above $\xi_j$ seems similar to 
the previous $U_{x}$ (Eq.~\eqref{eq:U1D}), 
but $\tilde{C}_2=C_2U_2$ gives an exact degenerate state in two dimensions
under the flux $\phi_L$.
The operator $U_x$ is basically the many-body polarization operator
~\cite{Resta1999}, and we will see later that
a many-body quadrupole operator~\cite{Wheeler2019,Kang2019,Ono2019,Tada2023} is also naturally introduced
in our approach.

Next, we consider the mirror symmetries $M_x,M_y$ about $x,y$-axis containing the origin.
They act as $M_x:(x_j,y_j)\to(x_j,L-y_j)$ and $M_y:(x_j,y_j)\to(L-x_j,y_j)$.
In contrast to $C_2$-rotation, mirror symmetries change the flux configuration from $+\phi_L$
to $-\phi_L$ in every plaquette.
The twisted mirror operators are given by
\begin{align}
\tilde{M}_x&=M_xR^x_{\pi}U_2,\quad \tilde{M}_{y}=M_yR^{x}_{\pi}.
\end{align}
One can easily see that these unitary operators commute with the Hamiltonian under the flux.

Now we can see that the Abelian group $C_{2v}=\{1,C_2,M_x,M_y\}$
leads to a projective representation in presence of the uniform flux $\phi_L$.
For example, 
\begin{align}
\tilde{M}_y\tilde{C}_2\tilde{M}_y^{-1}
&=C_2\cdot (M_yR^x_{\pi})U_2(M_yR^x_{\pi})^{-1} \nonumber\\
&=C_2\cdot \exp \left(-i\sum_j\xi_jS^z_{L-x_j,y_j}\right),
\end{align}
where $S^z_{x_j,y_j}=S^z_j$.
This is essentially the same 
exponetial factor as in the one-dimensional case. It has contributions from $y_j\neq0$ sites 
and for a fixed $y\neq 0$,
\begin{align}
-\sum_{x_j=0}^{L-1}\xi_{j}S^z_{L-x_j,y} 
=\sum_{x_j=0}^{L-1}\frac{2\pi}{L}x_{j}S^z_{x_j,y}-2\pi\sum_{x_j=1}^{L-1}S^z_{x_j,y}.
\label{eq:yC2_exp}
\end{align}
The last term is $2\pi\times (\mbox{odd number of } S^z_j)=\pi$ (mod $2\pi$)
for $y=1,2,\cdots,L-1$ with an even $L$, and the total factor is $-2\pi\sum_{x,y=1}^{L-1}S_{x,y}^z=\pi$
(mod $2\pi$).
Therefore, we obtain a non-trivial commutation relation
\begin{align}
\tilde{M}_y\tilde{C}_2=-\tilde{C}_2\tilde{M}_y.
\label{eq:sigmaC2}
\end{align}
With this non-trivial commutation relation, 
we can prove the LSM theorem for the two-dimensional system. 
Similarly, commutation relations with $\tilde{M}_x$ are also non-trivial and they are found to be
\begin{align}
\tilde{M}_x\tilde{C}_2=-\tilde{C}_2\tilde{M}_x, \quad \tilde{M}_x\tilde{M}_y=-\tilde{M}_y\tilde{M}_x.
\end{align}
Furthermore, squares of the twisted mirror operators are trivial, $(\tilde{M}_{x})^2=(\tilde{M}_{y})^2=1$,
but it is non-trivial for the twisted $C_2$-rotation,
$(\tilde{C}_2)^2=-1$, since $(\tilde{C}_2)^2 
=e^{i2\pi\sum_{x,y\neq0}S_j^z}=e^{i\pi}$ when $L$ is even.
These multiplication rules defines a projective symmetry of $C_{2v}=\{1,C_2,M_x,M_y\}$.
The factor system $\omega$ in the multiplication rule, $\tilde{g}_1\tilde{g}_2=\omega(g_1,g_2)\widetilde{g_1g_2},
(g_1,g_2\in C_{2v})$, is
summarized in Table~\ref{tb:Cayley}.
We see $\omega(g_1,g_2)/\omega(g_2,g_1)=-1$ for $g_1\neq g_2$.
\begin{table}[htbp]
  \centering
  \begin{tabular}{|c|c|c|c|}
    \hline\hline
    $\quad$ & $\quad\tilde{C}_2\quad$ & $\quad\tilde{M}_1\quad$ & $\quad\tilde{M}_2\quad$ \\ \hline
    $\quad\tilde{C}_2\quad$ & $-1$ & $+1$ & $-1$ \\ \hline
    $\tilde{M}_1$ & $-1$& $+1$ & $-1$ \\ \hline
    $\tilde{M}_2$ & $+1$ & $+1$ & $+1$ \\ \hline\hline
  \end{tabular}
  \caption{The factor system $\omega(g_1,g_2)$ for $\{\tilde{C}_2,\tilde{M}_1,\tilde{M}_2\}$.
The mirrors $\{\tilde{M}_1,\tilde{M}_2\}$ are either $\{\tilde{M}_x,\tilde{M}_y\}$ or 
$\{\tilde{M}_{xy},\tilde{M}_{\bar{xy}}\}$. The identity operator is omitted for simplicity.
 }
  \label{tb:Cayley}
\end{table}

Although Eq.~\eqref{eq:sigmaC2} is enough to prove the LSM,
we can also consider another Abelian subgroup $\{1,C_2,M_{xy},M_{\bar{xy}}\}$
containing the other two mirrors, $M_{xy}:(x_j,y_j)\to(y_j,x_j)$ and $M_{\bar{xy}}:(x_j,y_j)\to(L-y_j,L-x_j)$.
The twisted mirror operators are given by
\begin{align}
\tilde{M}_{xy}&=M_{xy}R^x_{\pi}U_{xy},\quad U_{xy}=e^{i2\pi/L^2\sum_{j}x_jy_jS_j^z},\\
\tilde{M}_{\bar{xy}}&=M_{\bar{xy}}R^x_{\pi}U_{\bar{xy}},\quad U_{\bar{xy}}=e^{-i2\pi/L^2\sum_{j}(L-x_j)y_jS_j^z}.
\end{align}
While the operator $U_2$ looks similar to the many-body polarization operator~\cite{Resta1999},
the present $U_{xy}$ operator may be regarded as a many-body quadrupole operator
~\cite{Wheeler2019,Kang2019,Ono2019,Tada2023}.
We find that
straightforward calculations for $\{\tilde{C}_2,\tilde{M}_{xy},\tilde{M}_{\bar{xy}}\}$
lead to the same commutation relations as those for $\{\tilde{C}_{2},\tilde{M}_x,\tilde{M}_y\}$.
Namely, $\tilde{C}_2\tilde{M}_{xy}=-\tilde{M}_{xy}\tilde{C}_2, 
\tilde{C}_2\tilde{M}_{\bar{xy}}=-\tilde{M}_{\bar{xy}}\tilde{C}_2, 
\tilde{M}_{xy}\tilde{M}_{\bar{xy}}=-\tilde{M}_{\bar{xy}}\tilde{M}_{xy}$
with $(\tilde{M}_{xy})^2=(\tilde{M}_{\bar{xy}})^2=1$.
This also defines a projective symmetry.

{\it Proof of Claim~\ref{thm2}.}
Armed with the above preparation,
we first prove exact degeneracy for the Hamiltonian $H(\phi_L)$ with the flux,
and then show that low energy spectra of $H(0)$ and $H(\phi_L)$ are essentially same in the thermodynamic limit.
For the proof of the former half, 
let us consider a partner state $\ket{\Phi}=\tilde{C}_2\ket{\Psi}$ where $\ket{\Psi}$
is a ground state of $H(\phi_L)$.
In the present case, the partner state $\ket{\Phi}$ is a ground state of $H(\phi_L)$ since 
the system has the $\tilde{C}_2$-symmetry, $[\tilde{C}_2,H(\phi_L)]=0$.
Furthermore, 
$\ket{\Phi}$ is orthogonal to the ground state $\ket{\Psi}$ because of 
the non-trivial commutation relation Eq.~\eqref{eq:sigmaC2}.
This proves the existence of pairs of the degenerate eigenstates for $H(\phi_L)$.

Next, we want to extrapolate the above results to the flux-free system.
To this end,
we consider the free energy for $H(\phi_L)$ which is an even function of $\phi_L$, 
since $H(0)$ has the $R^x_{\pi}$-symmetry which flips $+\phi_L\to-\phi_L$. 
As we have discussed, 
the flux per plaquette in a finite system of linear size $L$
is quantized in the unit of $\phi_L$ defined in Eq.~\eqref{eq:phi_L}.
Nevertheless, since $\phi_L \to 0$ as $L\to\infty$,
we can regard the flux per plaquette $\phi$ as a continuous variable in the thermodynamic limit.
We can also consider the free energy density $f$ in the thermodynamic limit, at an arbitrary temperature.
As in the case of the prominent St\v{r}eda formula~\cite{Streda1982,Streda1983,Widom1982,MacDonald1983},
it is natural to 
expect the free energy density to be a smooth function of the flux density, $f(\phi)$.
Then it can be Taylor expanded around $\phi=0$, as 
$f(\phi)=f_0+f_2\phi^2+f_4\phi^4+\cdots$,
where the coefficients $f_j$ are $O(1)$ constants.
This implies that the total free energy of the finite size system behaves as 
\begin{equation}
  F^{(L)}(\phi) = L^2 f(\phi) = F_0 + F_2 \phi^2 + F_4 \phi^4 + \cdots,
\end{equation}
where $F_j = f_j L^2$.
For the minimal flux~\eqref{eq:phi_L}, the total free energy is 
$F_0 + O(L^{-2})$ which is modified only slightly ($O(L^{-2})$) from the flux-free value $F_0$.
This implies that the ground-state degeneracy at zero flux is 
the same as the one under the minimal flux $\phi_L$

On the other hand, if the $R^x_{\pi}$-symmetry is spontaneously broken, the thermodynamic free energy density
should contain a non-analytic singular term such as $f_1|\phi|$ so that $\lim_{\phi\to-0}\partial f/\partial \phi
\neq \lim_{\phi\to+0}\partial f/\partial \phi$, and the above discussion does not apply.
In this case, however, there must be ground state degeneracy at $\phi=0$
corresponding to the $R^x_{\pi}$-symmetry breaking.
In any case, the ground states of $H(0)$ are not uniquely gapped in the thermodynamic limit, 
if the system with the flux $\phi_L$ has ground state degeneracy.
This completes the proof of the claim.
Note that the $R_{\pi}^x$-symmetry of $H(\phi=0)$ implies that there is no extra flux generated by
the magnetic coupling $J_{jk}$. The above argument is not applicable to a system with extra fluxes
which break the $R_{\pi}^x$-symmetry.

\subsection{Projective symmetry in different gauge}
\label{sec:simple2}
It is important to see that the commutation relations in the previous section are not specific to the
Landau gauge field configuration shown in Fig.~\ref{fig:string1} and are gauge independent.
Suppose that the two Hamiltonians $H(A)$ and $H(A')$ with different gauge fields $A, A'$ are related by a gauge transformation,
\begin{align}
H(A')={\mathcal U}H(A){\mathcal U}^{-1},\quad
{\mathcal U}=e^{i\sum_j\lambda_{j}S_j^z},
\end{align}
where $\lambda_j$ is a corresponding scalar function.
The two gauge fields are related as $A'=A+d\lambda$ and give the same flux 
$\sum_{\langle ij\rangle\in p}A_{ij}=\sum_{\langle ij\rangle\in p}A'_{ij}=\phi_p$ 
for every plaquette $p$.
When the Hamiltonian $H(A)$ has the symmetries $[\tilde{g}_i,H(A)]=0$ for a twisted operator $\tilde{g}_i$,
the corresponding symmetries of $H(A')$ are given by
\begin{align}
\tilde{g}_i'={\mathcal U}\tilde{g}_i{\mathcal U}^{-1}.
\label{eq:g_gauge}
\end{align}
This implies that,
when the commutation relation of $\tilde{g}_i,\tilde{g}_j$ is 
$\tilde{g}_i\tilde{g}_j=-\tilde{g}_j\tilde{g}_i$ for the gauge field $A$,
that for the other gauge field $A'$ is also $\tilde{g}_i'\tilde{g}_j'=-\tilde{g}_j'\tilde{g}_i'$.
Furthermore, the ground state expectation value of $\tilde{g}_i$ is also gauge independent,
because the ground states of $H(A)$ and $H(A')$ are related as $\ket{\Psi(A')}={\mathcal U}\ket{\Psi(A)}$.

\begin{figure}[tbh]
\includegraphics[width=8.0cm]{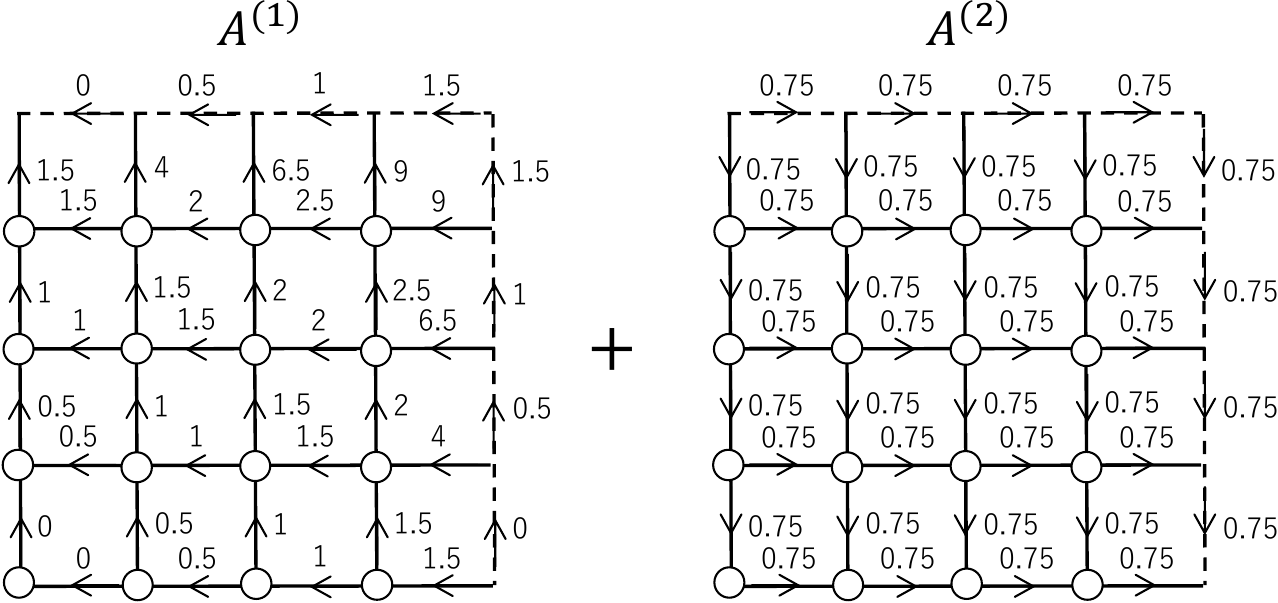}
\caption{The symmetric string gauge for a $L_x=L_y=L=4$ square lattice with
the periodic boundary condition.
Each number on the bonds corresponds to $A_{ij}=A_{ij}^{(1)}+A_{ij}^{(2)}$ in unit of $\phi_L=2\pi /L^2$.
}
\label{fig:string_sym}
\end{figure}

We consider the XXZ model Eq.~\eqref{eq:H2D0} with a different gauge field configuration and
explicitly check the gauge independence of 
the projective commutation relations.
Here, we focus on the $C_2$-rotation about the origin and 
$M_{xy}$-mirror about the $x=y$ line including the origin.
The new gauge field configuration is shown in Fig.~\ref{fig:string_sym}, where the two directions $x,y$ are treated on an equal footing
corresponding to a symmetric gauge $\vec{A}=B/2(-y,x,0)$ in the Euclidian space.
This symmetric string gauge is given by a sum of two terms $A=A^{(1)}+A^{(2)}$,
where $A^{(1)}$ corresponds to the symmetric gauge on the lattice torus
while $A^{(2)}$ does to a uniform flux piercing through the non-contractible loops of the torus.
The uniform part $A^{(2)}$ has been introduced so that the total fluxes through the non-contractible loops
are identical to those in the Landau string gauge.
Then, as will be discussed in Sec.~\ref{sec:extension}, 
the present symmetric string gauge can be obtained by a gauge transformation 
from the Landau string gauge.

The twisted $C_2$-rotation in the symmetric string gauge is found to be
\begin{align}
\tilde{C}_2'&=C_2U_2', \quad U_2'=e^{i\sum_j\xi_j' S_j^z},\\
\xi_j'&=\frac{\pi}{L}(x_j-y_j+L)(1-\delta_{x_j,0})(1-\delta_{y_j,0}).
\end{align}
Although $\xi_j'$ looks a bit complicated in the symmetric gauge,
the twisted $M_{xy}$-mirror takes a simple form,
\begin{align}
\tilde{M}_{xy}'&=M_{xy}R^x_{\pi}.
\end{align}
Then, the commutation relation is easily evaluated as
\begin{align}
\tilde{M}_{xy}'\tilde{C}_2'\tilde{M}_{xy}'^{-1}
=C_2\cdot \exp\left( -i\sum_j\xi_j'S_{y_j,x_j}^z\right),
\end{align}
where 
\begin{align}
-\sum_j\xi_j'S_{y_j,x_j}^z=\sum_j\xi_j'S_{x_j,y_j}^z-2\pi\sum_{x_j,y_j\neq0}S_j^z.
\label{eq:xyC2_exp}
\end{align}
The last term is $2\pi\times (\mbox{odd number of }S_j^z)=\pi$ (mod $2\pi$) similarly to the previous section.
Therefore, we obtain
\begin{align}
\tilde{M}_{xy}'\tilde{C}_2'=-\tilde{C}_2'\tilde{M}_{xy}'.
\label{eq:xyC2}
\end{align}
This is exactly the same as the previsouly obtained commutation relation in the Landau string gauge,
and other elements $\tilde{M}_x', \tilde{M}_y',\tilde{M}_{\bar{xy}}'$ can be discussed in a similar manner.
Therefore, we can explicitly see that the projective commutation relations 
are gauge independent, although the symmetry operators themselves are gauge dependent.
The new twisted operators can lead to the same LSM theorem as in the previous section.

\subsection{Time-reversal symmetry}
\label{sec:uniformTRS}
We have used the on-site ${\mathrm U}(1)\rtimes{\mathbb Z}_2$ symmetry 
and Abelian point group $C_{2v}$ in the previous section.
Here, we prove Claim~\ref{thm2TR} which is 
based on the time-reversal symmetry ${\mathbb Z}_2^T$ instead of the ${\mathbb Z}_2$ spin rotation symmetry.
Besides, spatial symmetry does not need to be the full $C_{2v}$ point group 
and the $C_2$-rotation symmetry alone is sufficient.
To be concrete, we again consider the Hamiltonian Eq.~\eqref{eq:H2DA} in the Landau string gauge.
We note that there is no flux (other than zero or $\pi$-flux) in the original Hamiltonian without the external flux since
the coupling $J_{jk}$ is real.

{\it Proof of Claim~\ref{thm2TR}}.
The proof of Claim~\ref{thm2TR} goes almost parallel to that of Claim~\ref{thm1TR}.
We focus on the two symmetries, namely the unitary $\tilde{C}_2$ symmetry and 
the anti-unitary time-reversal symmetry $T$.
Both of them are symmetries of the Hamiltonian $H(\phi_L)$ under the uniform flux and they commute each other
\begin{align}
\tilde{C}_2T=T\tilde{C}_2.
\label{eq:C2T}
\end{align}
The key observation is that the unitary $\tilde{C}_2$-operator satisfies $(\tilde{C}_2)^2=-1$
and corresponding eigenvalues are $c=\pm i$,
as discussed in the previous section (see Table~\ref{tb:Cayley}).
Let $\ket{\Psi}$ be a simultaneous eigenstate of $H(\phi_L)$ and $\tilde{C}_2$ with an eigenvalue $c$.
Then, the partner state $\ket{\Phi}=T\ket{\Psi}$ has the same energy because of the time-reversal symmetry
$[H(\phi_L),T]=0$, but has a different $\tilde{C}_2$-eigenvalue 
$c^{\ast}\neq c$.
Therefore, they are orthogonal. 
Note that the mirror symmetries $\tilde{M}$ cannot lead to an LSM theorem, beucase $\tilde{M}^2=+1$
and its eigenvalues are $\pm 1$ for an even $L$.
This is true even if we redefine the twisted operators as $\tilde{g}\to i\tilde{g}$,
because the commutation relation with $T$ also changes accordingly.

The next step is to extrapolate the above result to the flux-free system
and obtain the LSM theorem for $H(\phi=0)$.
In the present case, 
the ${\mathbb Z}_2$ spin rotation symmetry is an extra symmetry and
an argument without use of this symmetry is desirable.
Naively, one would expect that the tiny flux $\phi_L=2\pi/L^2$ could have negligibly small
effects on the energy spectrum, although it is non-trivial since the total flux in the entire system is $O(1)$.
Here, we provide an argument based on this intuition 
and also on topological triviality of the real magnetic coupling $J_{jk}$.
We map the spin-$S$ XXZ model onto a hard-core boson model $H_{B}$~\cite{Sachdev}.
The Hamiltonian $H_B$ consists of a kinetic term with the flux $e^{iA_{jk}}S^+_jS^-_k\sim e^{iA_{jk}}b^{\dagger}_jb_k$ 
and a density-density interaction $S^z_jS^z_k\sim (n_j-\nu)(n_k-\nu)$, 
where $b_j$ is the annihilation operator of the boson.
$n_j=b^{\dagger}_jb_j$ is the density operator and $\nu=S$ is the filling per site.
The hard-core condition is described by an on-site interaction, $H_U=U\sum_j \prod_{m=0}^{2S}(n_j-m)$
with $U\to\infty$.
Note that the Chern numbers of 
the single-particle bands of the bosons at $\phi=0$ are zero, 
since the hopping integrals $J_{jk}$
are real. 
At the limit $U=0$ and $J^z_{jk}=0$, the many-body spectrum of $H_B$ 
is almost unchaged by the flux,
because the change of each Landau level is vanishingly small as was discussed previously~\cite{Tada2021,Berkolaiko2013} (see Appendix~\ref{app:stability}).
Similarly,
the energy spectrum in the interacting case is also robust to the flux, essentially because the density-density
interactions do not include any flux~\cite{Tada2021}.
Therefore, we conclude that the energy spectra of the XXZ Hamiltonians $H(\phi_L)$ and $H(0)$ are essentially
same in the thermodynamic limit.
This completes the proof of Claim~\ref{thm2TR}.
We note that our argument is not applicable to a Chern band system with fluxes generated by (complex)
magnetic coupling $J_{jk}$. 
In such a system, Landau level degeneracy will change under an external flux and consequently 
degeneracy of the many-body spectrum could also change~\cite{Streda1982,Streda1983,Widom1982,MacDonald1983,Higashino2025}. 
Similarly, fractional quantum Hall systems and fractional Chern insulators which are built on Landau levels/Chern bands
are also outside the scope of the present study.
On the other hand, the above argument holds for a $\pi$-flux generated by positive and negative $J_{jk}$,
since there is no spin current anywhere in the system with the Hamiltonian $H(0)$ (Appendix~\ref{app:stability}).

\subsection{Extensions}
\label{sec:extension}
The above discussions can be extended to a large class of models.
Firstly, we consider real magnetic interactions $J_{ij}$ whose ranges are longer than the nearest neighbors,
and discuss gauge fields which give a small uniform flux.
Secondly, we examine the gauge transformation appearing in the projective symmetry operators
from a general point of view.
Based on these observations, we finally discuss generalization of our argument for the square lattice
to other lattices, which completes the proof of the LSM theorems.

{\it Further neighbor interactions.}
We explicitly consider further neighbor magnetic interactions $J_{jk}$ with the on-site U(1)$\rtimes{\mathbb Z}_2$ 
up to third nearest neighbors.
Here, we focus on the Landau string gauge.
Although it seems a priori non-trivial how to define a gauge field configuration for a system with general
interactions, it is easy to find a concrete configuration which consistently realize the uniform flux 
for general plaquettes connected with non-zero $J_{jk}$.
For example, we consider the second nearest neighbor interaction and third nearest neighbor interaction.
\begin{figure}[tbh]
\includegraphics[width=8.0cm]{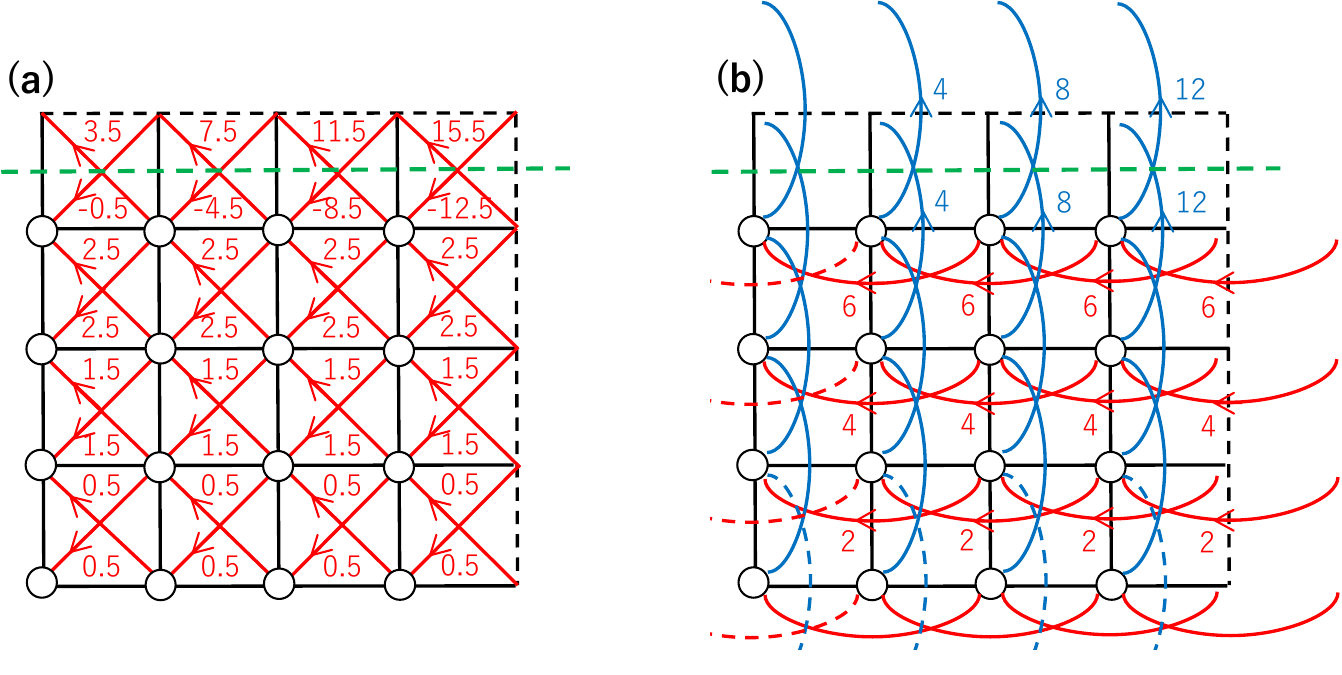}
\caption{The Landau string gauge for (a) the second nearest neighbor and (b) the third nearest neighbor interactions.
Colors are used for the eyes.
Each number on the bonds corresponds to $A_{ij}$ in unit of $\phi_L=2\pi /L^2$ with $L=4$.
}
\label{fig:string2}
\end{figure}
The gauge field for the latter coupling is shown in Fig.~\ref{fig:string2} (a), where the flux for a triangular plaquette
$p$ whose area is half of a square plaquette satisfies $\phi_p=\phi_L/2$.
Similarly,
a gauge field for a third nearest neighbor interaction is obtained straightforwardly as shown in Fig.~\ref{fig:string2} (b).
One can see that the flux for a general plaquette $p$ with the area $|p|$ is given by 
$\phi_p=\phi_L|p|$ as naturally expected.
Note that the flux for a rectangular plaquette is given in several ways and they give the same flux
consistently. 

In general, a consistent gauge field configuration is described by $A_{jk}=A_{jk}^x+A_{jk}^y$,
\begin{align}
A_{jk}^x&=-\phi_L\frac{y_j+y_k}{2}(x_j-x_k),
\label{eq:Ax} \\
A_{jk}^y&=\phi_L\frac{x_j+x_k}{2}L\delta_{\langle jk\rangle},
\label{eq:Ay}
\end{align}
where $\delta_{\langle ij\rangle}=+1 (-1)$ when the link $\langle ij\rangle$ goes through the seam in the $+y$-direction
($-y$-direction)
and  $\delta_{\langle ij\rangle}=0$ otherwise.
In the above definition, the $y$-coordinate is given by the torus coordinate $y_j=0,1,\cdots,L-1$,
while the $x$-coordinate is described by the non-periodic coordinate $x_j=\cdots,0,1,\cdots,L-1,L,\cdots$.
We can show that this gauge field indeed gives the uniform flux in a consistent manner as discussed
in Appendix~\ref{app:flux}.

It should be noted that 
the above arguments are essentially applicable to the symmetric string gauge as well.
This is because the Landau string gauge and symmetric string gauge 
have the same loop integrals and hence are connected by a gauge transformation
as will be discussed in the following..

{\it Existence of gauge transformation.}
It might be non-trivial whether or not there exists a gauge transformation which realizes
projective symmetries such as the $\tilde{C}_2$-rotation for a given set of $J_{jk}$ and $A_{jk}$.
Here, we claim that such a gauge transformation does exist because all the fluxes including
the ``global" fluxes through
non-contractible loops on the torus are conserved by the point group symmetries. 
To understand this, let us first consider a continuum system on a torus and
the problem is whether or not there is a gauge transformation between 
given two gauge fields $A$ and $A'$.
A gauge transformation exist if $a\equiv A'-A$ is an exact 1-form, $a=d\xi$. 
Although 
there are non-contractible loops on a torus,
we have a well-known necessary and sufficient condition for a 1-form to be an exact form on a plane
with holes;
{\it a 1-form $\omega$ is exact if and only if
any closed loop integral of $\omega$ is vanishing}~\cite{Nakahara}. 
This statement holds also on a lattice and up to modulo $2\pi$ when a loop integral is vanishing in modulo $2\pi$.

For example for the $C_2$-rotation, 
the two gauge fields $A_{jk}$ and $A'_{jk}=A_{C_2j,C_2k}$ in the Landau string gauge
clearly give an identical uniform flux for any plaquette $p$,
$\phi_p=\sum_{\langle ij\rangle\in p}A_{ij}=\sum_{\langle ij\rangle\in p}A'_{ij}$,
namely $\sum_{\langle ij\rangle\in p}a_{ij}=0$.
Besides, we can see that the integral of $a=A'-A$ along a non-contractible loop is vanishing (in modulo $2\pi$).
For a loop $C_x$ parallel to the $x$-axis with a fixed $y$-position,
the flux is 
\begin{align}
\Phi_x(y)&=\sum_{\langle jk\rangle \in C_x}a_{jk}\nonumber\\
&=\sum_{\langle jk\rangle \in C_x}-\phi_L(L-y)(-x_j+x_k)+\phi_Ly(x_j-x_k)\nonumber\\
&=\phi_LL^2= 0 \quad (\mbox{mod }2\pi).
\end{align}
Similarly for a loop $C_y$ parallel to the $y$-axis with a fixed $x$-position,
\begin{align}
\Phi_y(x)&=\sum_{\langle jk\rangle \in C_y}a_{jk}\nonumber\\
&=\sum_{\langle jk\rangle \in C_y}\phi_L(L-x)L\delta_{\langle kj\rangle}-\phi_LxL\delta_{\langle jk\rangle}\nonumber\\
&=-\phi_LL^2= 0 \quad (\mbox{mod }2\pi).
\end{align}
A general non-contractible loop which is not necessarily parallel to $x$- or $y$-direcion
is decomposed into a sum of the above $C_x$ or $C_y$ and contractible plaquette loops.
In this way, any closed loop integral of $a=A'-A$ is vanishing  in modulo $2\pi$ on the torus,
and thus there exists a scalar function $\xi_j$ such that $a_{jk}=d\xi_{jk}=\xi_j-\xi_k$ corresponding to
the gauge transformation. 
(This is in contrast to the magnetic translation symmetry with $\phi_L=2\pi/L^2$ 
where there is no gauge transformation for the one site translation~\cite{Tada2021}.)
Similar arguments apply also to the mirror symmetries as well.
Besides, 
once a set of projective symmetry operators has been obtained for the Landau string gauge,
they can be transformed to a different gauge field configuration such as the symmetric string gauge,
as seen in Eq.~\eqref{eq:g_gauge}

\begin{figure}[tbh]
\includegraphics[width=8.0cm]{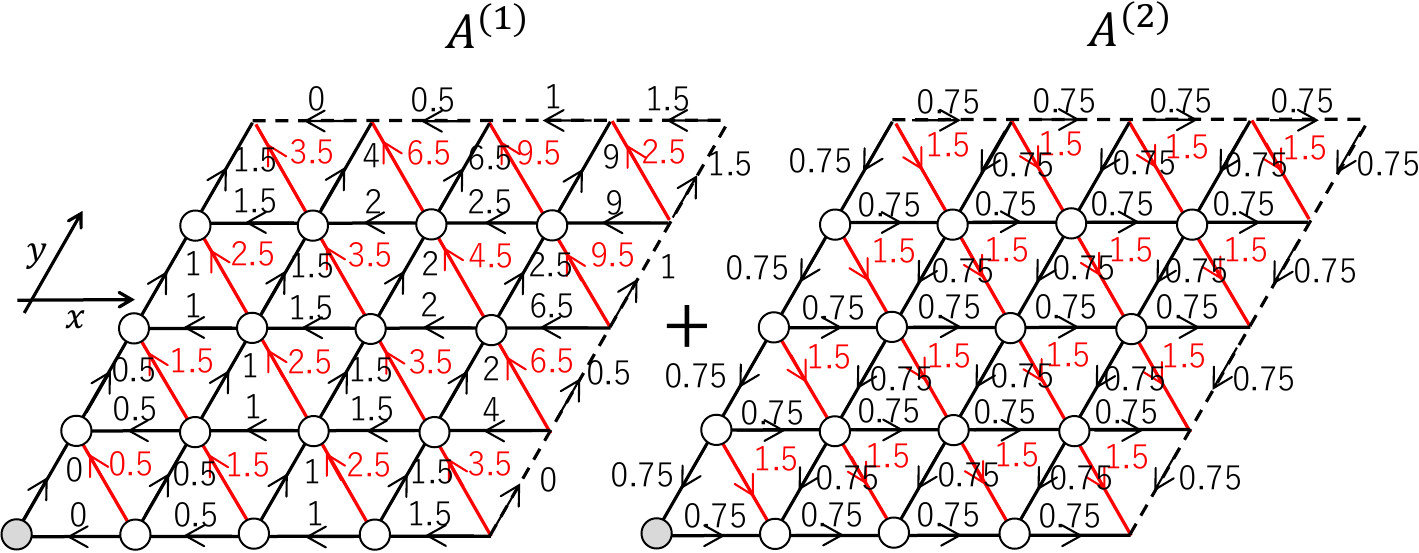}
\caption{The symmetric string gauge $A=A^{(1)}+A^{(2)}$ (in unit of $2\pi/4^2$)
for the triangular lattice with the nearest neighbor interaction, where
$A^{(2)}$ is a uniform gauge field.
Colors are used for the eyes.
The $x$- and $y$-axes are taken to be parallel to the primitive lattice vectors.
}
\label{fig:triangularlatt}
\end{figure}

{\it Other lattices.}
Based on the above argument of general coupling $J_{jk}$,
one can easily generalize the LSM theorem for the square lattice to other lattices 
in two dimensions.
Indeed, essentially the same proof  applies for 
(face-centered) rectangular and hexagonal lattices 
with a site-centered $C_2$-rotation and mirror symmetries.
For example, the LSM theorem holds for a triangular lattice.
The corresponding 
symmetric string gauge for the nearest neighbor
interaction is shown in Fig.~\ref{fig:triangularlatt}, where the system has the $\tilde{C}_2$-rotation
about the origin and $\tilde{M}_{xy},\tilde{M}_{\bar{xy}}$ about the $x=\pm y$ lines containing the origin.

\begin{figure}[tbh]
\includegraphics[width=8.0cm]{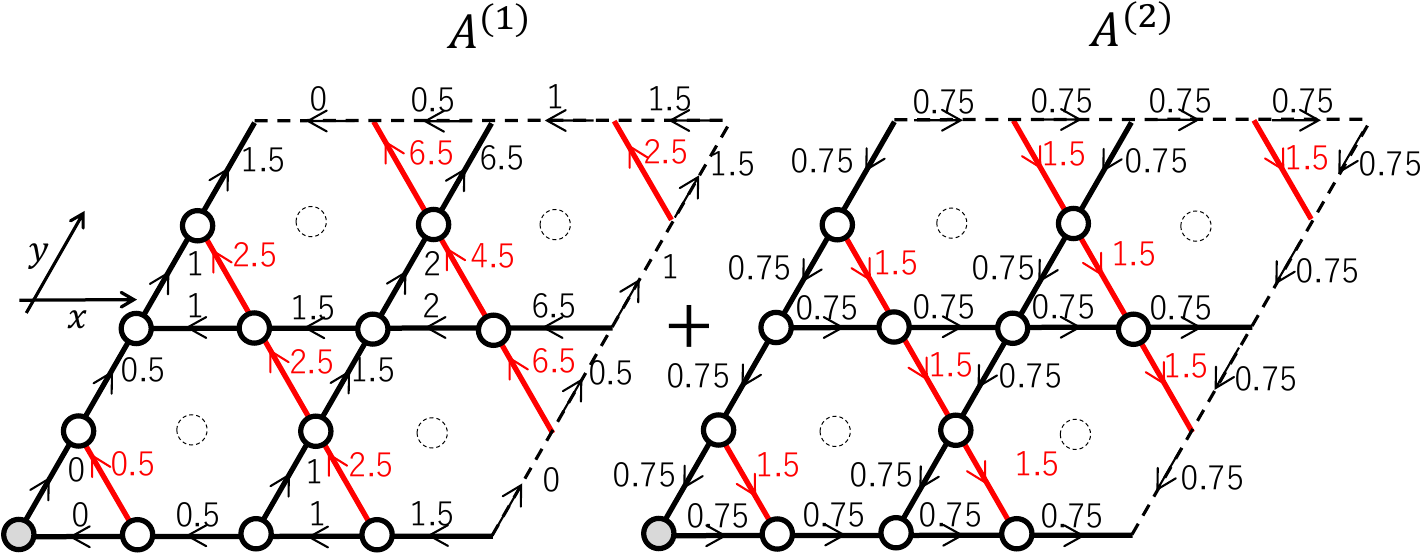}
\caption{The symmetric string gauge 
for the Kagome lattice with the nearest neighbor interaction (in unit of $2\pi/4^2$ corresponding to the total flux $2\pi$).
The thin dotted circles are the sites depleted from the triangular lattice.
Colors are used for the eyes.
}
\label{fig:kagome}
\end{figure}

Similarly,
our argument holds also for Kagome lattice with $C_{2v}$ point group symmetry around a site. 
The Kagome lattice is constructed by depleting sites from the triangular lattice as shown in Fig.~\ref{fig:kagome}.
In this lattice, the factor in Eq.~\eqref{eq:yC2_exp} is 
$2\pi\times(\mbox{odd number of }S_j^z)=\pi$ (mod $2\pi$)
and leads to the non-trivial commutation relation Eq.~\eqref{eq:sigmaC2}.
The $\tilde{C}_2$ operator is squared to $-1$ and the commutation relation Eq.~\eqref{eq:C2T} also holds.
Therefore, 
the point group LSM theorems hold in the Kagome lattice as well, which is consistent with the previous LSM theorem
by the translation symmetry~\cite{Oshikawa2000,Hastings2004}.

\begin{figure}[tbh]
\includegraphics[width=8cm]{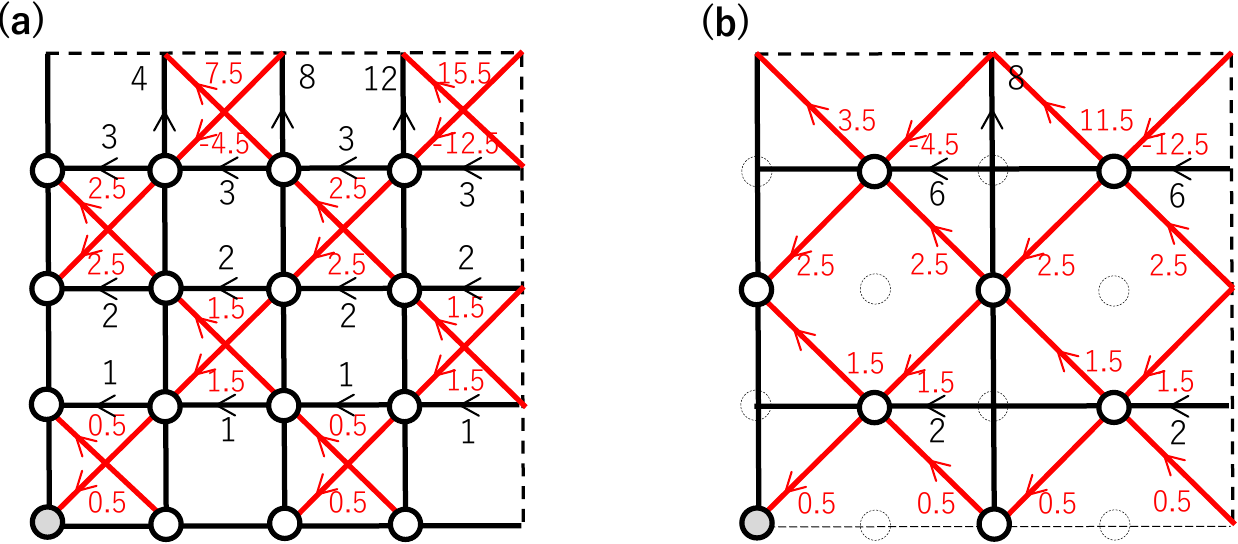}
\caption{The Landau string gauge 
for the checkerboard lattice (in unit of $2\pi/4^2$ corresponding to the total flux $2\pi$).
(a) The primitive translation vectors are $\pi/4$-rotated from the original $x,y$-axes  and (b) they are parallel
to the $x,y$-axes.
Depleted sites are shown by the thin dotted circles.
Colors are used for the eyes.
}
\label{fig:checkerboard}
\end{figure}

We can also consider the checkerboard model which is obtained by
adding the next nearest neighbor interaction to the square lattice model (Fig.~\ref{fig:checkerboard} (a)).
(To make the primitive translation vectors parallel to the original $x,y$-axes,
one can deplete sites in an appropriate way  as shown in Fig.~\ref{fig:checkerboard} (b)).
It is easy to see that the factor in Eq.~\eqref{eq:yC2_exp} 
and the commutaion relation of the combined operators are non-trivial,
which leads to ground state degeneracy. 
(Similarly, one can also see that 
the ground state degeneracy can be derived from $(\tilde{C}_2)^2=-1$ and Eq.~\eqref{eq:C2T}.)
Indeed, it was shown that the $S=1/2$ checkerboard XXZ model has
the plaquette singlet ground state with two fold degeneracy~\cite{Fouet2003,Shannon2004}.
The ground states break the $C_2$-rotation symmetry and if present, 
the translation symmetry simultaneously.
We stress that,
in contrast to the other examples discussed above, 
the ground state degeneracy of the checkerboard model is not obtained from the translation LSM theorem 
because there are two half-integer spins in the unit cell.
In the present system, the point group symmetry can lead to a stronger statement than that by 
the translation symmetry alone,
and it is consistent with the improved LSM theorem by the translation and inversion symmetries~\cite{Furuya2019}.
Note that the ground state degeneracy of the checkerboard model cannot be imposed solely by the time-reversal
symmetry, because the total number of spins is always even under the periodic boundary condition.

\begin{figure}[tbh]
\includegraphics[width=5.0cm]{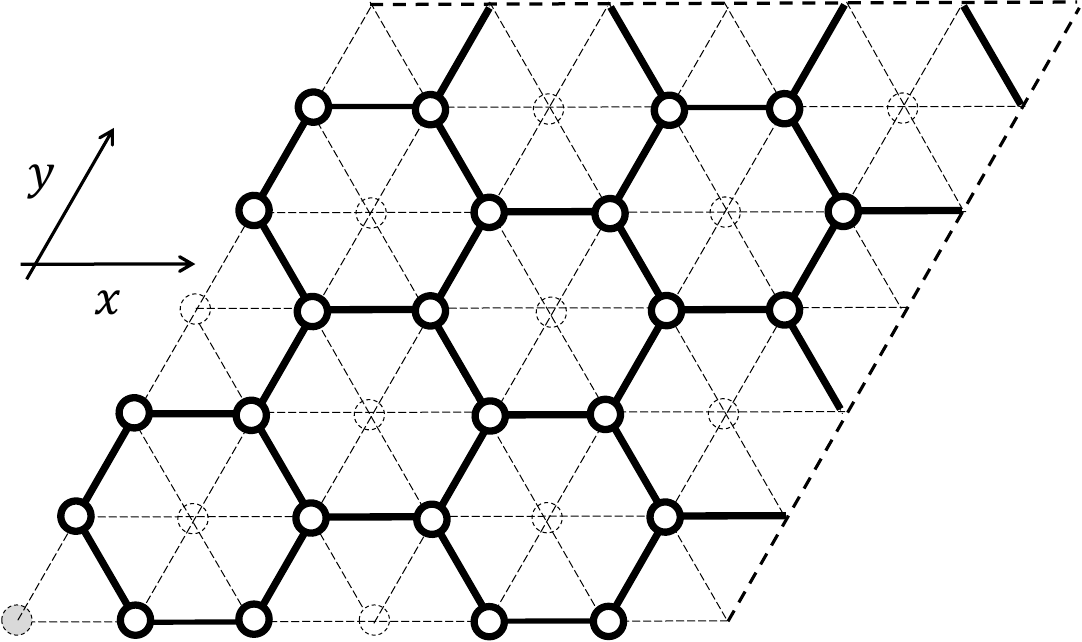}
\caption{The honeycomb lattice with the ``armchair type" periodic boundary condition.
The gauge field is the same as that in Fig.~\ref{fig:triangularlatt} (in unit of $2\pi/6^2$) 
which gives the total flux $2\pi$ for the entire
system.
The thin dashed bonds and sites are those of the triangular lattice and are shown for the eyes.
}
\label{fig:honeycomb}
\end{figure}

However, our discussion is not applicable to the honeycomb lattice. 
Let us consider the honeycomb lattice with the ``armchair type" periodic boundary condition,
where sites corresponding to the hexagonal plaquatte centers
are depleted from the triangular lattice
as shown in Fig.~\ref{fig:honeycomb}. 
We can define the point group symmetry $C_{2v}$ around the origin (which is a depleted site),
and just repeat the same argument as before.
Then, we find that the extra factor in Eq.~\eqref{eq:xyC2_exp} 
is $2\pi\times(\mbox{even number of }S_j^z)=0$ (mod $2\pi$)
and cannot obtain a non-trivial commutation relation, which does not lead to ground state degeneracy. 
This is consistent with the previous studies where a unique gapped ground state
has been constructed for $S=1/2$ spins~\cite{Metlitski2018,Kimchi2013}.
We note that ground states can be degenerate under the open boundary condition.
For example, a system consisting of three nearby hexagons has 13 (odd number) spins, 
where the $\mathbb{Z}_2\times\mathbb{Z}_2$ spin rotation symmetry or the time-reversal symmetry leads to 
ground state degeneracy.
This would imply that internal symmetry alone is insufficient for robust low energy spectrum.

A summary for the above discussed lattices is given in Table~\ref{tb:lattices}.
Generalizing these arguments for the representative lattices, 
we can prove the point group LSM theorem for the half-intger
spin XXZ model on general two-dimensional
lattices with $C_{2v}$ about a site with a spin, as discussed in the following. 
\begin{table}[htbp]
  \centering
  \begin{tabular}{|c|c|c|c|c|c|}
    \hline\hline
    $\quad$& square & triangular & Kagome & checkerboard & honeycomb\\ \hline
    PG & Yes & Yes & Yes & Yes & No \\ \hline
    TL & Yes & Yes & Yes & No & No \\ \hline\hline
  \end{tabular}
  \caption{Summary for the representative lattices. ``Yes" (``No") means that the LSM theorem
is (not) applicable to the half-integer spin XXZ model on that lattice. 
PG (TL) represents the LSM theorem by the point group symmetry (translation symmetry).}
  \label{tb:lattices}
\end{table}

{\it Proof of claim for general two-dimensional lattices}.
First, the commutation relation of the combined symmetry operators are characterized by
$\exp(2\pi\sum_{x,y\neq0}S_j^z)=\exp(i\pi |\Lambda_0|)$, where $|\Lambda_0|$ is the number of
lattice sites with $x,y\neq0$.
By denoting the number of lattice sites at the two lines $\Lambda_{0x}=\{(x,y)|x=0\}$ and 
$\Lambda_{0y}=\{(x,y)|y=0\}$, we have
$|\Lambda_0|=|\Lambda|-|\Lambda_{0x}|-|\Lambda_{0y}|+1$ where ``$+1$" corresponds to the contribution
of the origin $(0,0)$.
We suppose that a half-integer spin is located at the origin. 
Because $|\Lambda_{0x}|=|\Lambda_{0y}|$ thanks to the mirror symmetry
in isotropic systems discussed in this study,
$\exp(i\pi |\Lambda_0|)=-1$ and the non-trivial commutation relations hold,
which leads to ground state degeneracy.
This completes our proof of the LSM theorem by the point group symmetry for general two-dimensional lattices.
We can further generalize the theorem to systems where integer spins and half-integer spins coexist
by repeating almost the same argument (Appendix~\ref{app:non-uniform}).

\section{Summary}
\label{sec:summary}

In this study, we have discussed the LSM theorems in half-integer spin systems
with the on-site ${\mathrm U}(1)\rtimes{\mathbb Z}_2$ spin-rotation symmetry and the $C_{2v}$ point group symmetry.
The LSM theorems with use of the ${\mathbb Z}_2^T$ time-reversal symmetry was also examined.
The statement is that the ground state of such a system cannot be uniquely gapped
under the periodic boundary condition.
We introduced a tiny uniform U(1) flux $\phi_L=2\pi/L^2$, 
under which the projective representation of the point group is constructed.
This enabled us to obtain degeneracy of eigenstates under the flux $\phi_L$, 
and the extrapolation down to zero flux is well-controlled since $\phi_L=O(L^{-2})$ has negligibly
small effects on energy spectra in two or three dimensions.
We obtained
similar degeneracy for a system with a local flux, and it could be extraplated to the flux-free system.
Technically, one of the important points is that there is a half-integer spin at the fixed point of 
the point group symmetry in both approaches, which leads to the non-trivial phase factor in the commutation relations.
Our approach is pedagogical and applicable to a wide class of systems including bosonic particle systems 
with particle-hole symmetry, 
although it essentially relies on the continuous $\mathrm{U}(1)\rtimes\mathbb{Z}_2$ 
or $\mathrm{U}(1)\times\mathbb{Z}_2^T$ symmetry
rather than the discrete $\mathbb{Z}_2\times\mathbb{Z}_2$ or $\mathbb{Z}^T$ symmetry. 
It could provide a useful perspective on roles of point group symmetries in quantum many-body systems.

\begin{acknowledgments}
We are grateful to Ken Shiozaki, Yuan Yao, and Yuan-Ming Lu for valuable discussions.
This work was supported by JSPS KAKENHI Grant Nos.
JP17K14333, JP22K03513, JP19H01808, JP23H01094, and JP24H00946,
and by JST CREST Grant No. JPMJCR19T2.
\end{acknowledgments}

\appendix

\section{Proof of Claim~\ref{thm1}}
\label{app:proof1}
Here, we do not introduce any flux.
To prove Claim~\ref{thm1}~\cite{LSM1961,AffleckLieb1986}, 
we consider a variational state $\ket{\Phi}=U_{x}\ket{\Psi}$ with respect to a ground state
$\ket{\Psi}$, where the LSM twist operator is given by Eq.~\eqref{eq:U1D}.
The variational energy is close to the ground state energy, $\bra{\Phi}H\ket{\Phi}-\bra{\Psi}H\ket{\Psi}=O(L^{-1})$.
Besides, the variational state is orthogonal to the ground state, $\bra{\Psi}\Phi\rangle=0$,
which is derived from a combined symmetry of the ${\mathbb Z}_2$ spin rotation and
inversion symmetries,
\begin{align}
\tilde{I}=I R^x_{\pi}.
\end{align}
This operater commutes with the Hamiltonian, $[\tilde{I},H]=0$, and is squared to unity,
$\tilde{I}^2=1$, for an even $L$. 
The $\tilde{I}$-operator is regarded as an inversion symmetry twisted by the on-site ${\mathbb Z}_2$ symmetry.
The introduction of the twisted symmetry $\tilde{I}$ is essential in the proof, 
although each of $I$ and $R^y_{\pi}$ already commutes with the Hamiltonian.
Now the ground state $\ket{\Psi}$ is a simultaneous eigenstate of $H$ and $\tilde{I}$.
The two unitary operators $U_{x}$ and $\tilde{I}$ have a non-trivial commutation relation,
\begin{align}
\tilde{I} U_{x}\tilde{I}^{-1}&=\exp\left( -i\frac{2\pi}{L}\sum_jx_jS_{L-j}^z\right),
\end{align}
where
\begin{align}
-\frac{2\pi}{L}\sum_{j=0}^{L-1}x_{j}S^z_{L-j} 
=\frac{2\pi}{L}\sum_{j=0}^{L-1}x_jS^z_{j}-2\pi\sum_{j=1}^{L-1}S^z_{j}.
\end{align}
The last term is $2\pi\times(\mbox{odd number of }S^z_j)=\pi$ (mod $2\pi$) since $L$ is even.
Therefore, we obtain the non-trivial commutation relation,
\begin{align}
\tilde{I} U_{x}=-U_{x}\tilde{I},
\label{eq:sigmaU1D}
\end{align}
for the present half-integer spin model.
Orthogonality of the two states  immediately follows from Eq.~\eqref{eq:sigmaU1D},
and indeed, $\langle \Psi\ket{\Phi}=\bra{\Psi}U_{x}\ket{\Psi}
=-\bra{\Psi}\tilde{I}U_{x}\tilde{I}^{-1}\ket{\Psi}=-\langle \Psi\ket{\Phi}=0$.
This means that the variational state $\ket{\Phi}$ is either a low energy excited state
or one of the degenerate ground states other than $\ket{\Psi}$ in the thermodynamic limit, and therefore 
a unique gapped ground state is not realized in the XXZ model.
Clearly, this proof can be generalized to a wide class of one-dimensional spin models
with the on-site ${\mathrm U}(1)\rtimes{\mathbb Z}_2$ and inversion symmetry.

\section{Stability of single-particle eigenenergies to external fluxes}
\label{app:stability}
We briefly discuss the stability of eigenenergies of the single-particle ``hopping Hamiltonian" $J_{jk}$ to 
an external flux~\cite{Tada2021,Berkolaiko2013}.
For simplicity, we consider the nearest neighbor $J_{jk}$ on the square lattice under the periodic boundary condition.
The hopping Hamiltonian $J$ is a real symmetric matrix, $J_{jk}=J_{kj}$, and
hence eigenvectors (single-particle wavefunctions) $\{\psi_n\}$ can be real.
We suppose that any eigenvalue $\varepsilon_n$ is non-degenerate, and if necessary, we will add a small perturbation
to remove degeneracy (the perturbation will be taken to be zero at the end of the calculation).

We consider the matrix $(J_A)_{jk}=e^{iA_{jk}}J_{jk}$, where the flux for a cycle $c$ is given by
$\theta_c=\sum_{\langle jk\rangle\in c}A_{jk}$.
In the present model, there are $b=|\mathcal{E}|-|\mathcal{V}|+1=2L^2-L^2+1=L^2+1$ independent cycles including 
the non-contractible loops around the torus, when the sites $\mathcal{V}$ and bonds $\mathcal{E}$ 
are regarded as a graph $G=(\mathcal{V},\mathcal{E})$.
Note that $\vec{\theta}=(\theta_1,\cdots,\theta_{b})$ is continuous, although a uniform flux for plaquettes 
can be realized only when $\theta_c$ is an integer multiple of $\phi_L=2\pi/L^2$. 
We fix the gauge configuration so that $A_{jk}\neq0$ only for a set of $b$ bonds, $S$.
The bonds $S$ is chosen so that the subgraph $G\backslash S$ is a spanning tree of the graph $G$.
The eigenenergy $\varepsilon_n(\vec{\theta})$ is differentiable since it is non-degenerate at $\vec{\theta}=0$.
One easily obtains $\partial \varepsilon_n(0)/\partial \theta_c=\braket{\psi_n|\partial J(0)/\partial \theta_c| \psi_n}
=\sum_{\braket{j_ck_c}\in S}{\rm Im}(\psi_{nj_c}^*\psi_{nk_c})=0$ since $\psi_n$ is real, where $J$ is the single-particle hopping Hamiltonian.
Therefore, $\nabla \varepsilon_n(\vec{\theta})=0$ at the point $\vec{\theta}=0$ in the flux space,
which implies that $\varepsilon_n(\vec{\theta})-\varepsilon_n(0)=O(\theta_c^2)$.  
When the Chern number for the corresponding band is zero, no spectral flow takes place between bands and thus 
$\varepsilon(\vec{\theta})$ will be smooth up to $\theta_c=\phi_L$.
This results in a vanishingly small energy change, $\varepsilon_n(\phi_L)-\varepsilon_n(0)=O(\phi_L^2)=O(L^{-4})$,
under the uniform flux $\phi_L$ for plaquettes.
Note that stability of a many-body energy to the flux follows from that of the single-particle energies,
basically because the many-body energy is given by an $O(L^2)$ sum of $\{\varepsilon_n\}$ and its change is at most $O(L^2)\times O(L^{-4})=O(L^{-2})$~\cite{Tada2021}.

Physically, stability of a many-body energy to the magnetic field $\phi_L$ may be almost trivial 
in a system where the Chern number is zero and the energy density behaves as
$\epsilon(\phi)=\epsilon(0)+a_n\phi^n+\cdots$ with $n>1$ in the thermodynamic limit.
In such a system with a finite size $L$, the energy change by the magnetic field $\phi_L=O(L^{-2})$ will be at most $O(L^{2})\times O(L^{-2n})
=O(L^{-2(n-1)})$ which vanishes in the thermodynamic limit.

\section{Flux for gauge field Eqs.~\eqref{eq:Ax},~\eqref{eq:Ay}}
\label{app:flux}
We demonstrate that the gauge field Eqs.~\eqref{eq:Ax},~\eqref{eq:Ay} consistently give
a uniform flux.
We first consider a triangular plaquette $p$ with the vertex positions $R_j=(X_j,Y_j), j=0,1,2$.
Note that $X_j$ is non-periodic while $Y_j$ has the periodicity $L$ in the definition of 
the gauge field Eq.~\eqref{eq:Ax},~\eqref{eq:Ay}.
We assume that distance between any of the three vetices is much smaller than the system size
and thus the triangle is uniquely defined on a torus with the periodic boundary condition.
\begin{figure}[tbh]
\includegraphics[width=8.0cm]{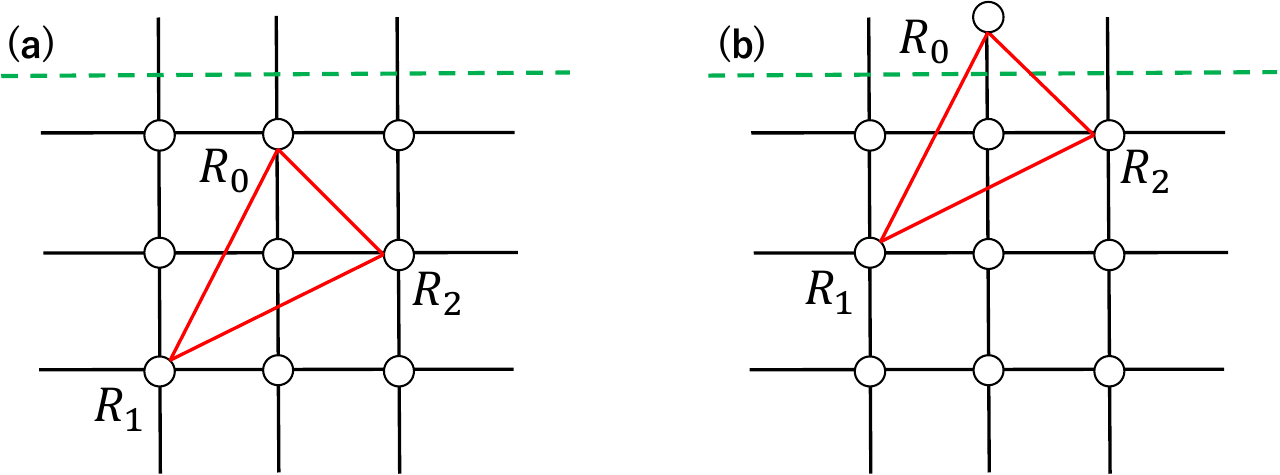}
\caption{Triangles with the vertices $R_0,R_1,R_2$. (a) No link between the vertices crosses the seam 
(green dashed line) and (b) two links cross it.
}
\label{fig:triangles}
\end{figure}
When any link between the vertices does not cross the seam and thus the periodicity of $y$
is not important as shown in Fig.~\ref{fig:triangles} (a), the flux for $p$ is given by
\begin{align}
\phi_p&=A^x_{R_0,R_1}+A^x_{R_1,R_2}+A^x_{R_2,R_0}\nonumber \\
&=\frac{\phi_L}{2}[(X_1-X_0)(Y_2-Y_0)-(Y_1-Y_0)(X_2-X_0)]\nonumber\\
&=\frac{\phi_L}{2}[(R_1-R_0)\times(R_2-R_0)]_z.
\end{align}
Note that $\phi_p$ is well-defined for the given triangle and does not change even if one takes into account
intermediate points between the vertices $R_0,R_1,R_2$,
because $A_{R_0,R_1}^x=A_{R_0,R'}^x+A_{R',R_2}^x$ on the line between $R_0,R_1$ and so on.
When two links cross the seam, we have contributions from $A^y_{jk}$ and the periodicity of $y$ plays a role.
Without loss of generality,
let us consider the case where the links $\langle R_0R_1\rangle$ and $\langle R_2R_0\rangle$ cross the seam
as shown in Fig.~\ref{fig:triangles} (b).
Then the flux is
\begin{align}
\phi_p&=(A^x_{R_0,R_1}+A^y_{R_0,R_1})+A^x_{R_1,R_2}+(A^x_{R_2,R_0}+A^y_{R_2,R_0})\nonumber \\
&=\frac{\phi_L}{2}[(R_1-R_0)\times(R_2-R_0)]_z+\frac{\phi_L}{2}(X_2-X_1)L\nonumber\\
&=\frac{\phi_L}{2}[(R_1-R_0')\times(R_2-R_0')]_z,
\end{align}
where $R_0'=R_0+(0,L)$ is the non-periodic coordinate.
Therefore, the flux is given by $\phi_p=\pm\phi_L|p|$ depending on the orientation of the triangle,
where $|p|$ is the area of the triangular plaquette.

For a general polygon plaquette $p$, 
we decompose it into a sum of triangles and apply the same argument for every
triangular plaquette.
Then, the flux  for the polygon plaquette is $\phi_p=\pm \phi_L|p|$.

In addition, loop integrals of $A_{ij}$ along non-contractible loops are also well-defined quantities.
Let us consider a non-contractible loop $C_x$ parallel to the $x$-axis at a fixed $y$-position,
\begin{align}
\Phi_x(A;y)&=\sum_{\langle jk\rangle\in C_x}A_{jk.}
\label{eq:PhiAx}
\end{align}
Eq.~\eqref{eq:PhiAx} is independent of a choice of a sequence $\{A_{jk}\}$
and is well-defined.
A similar argument applies to a loop $C_y$ parallel to the $y$-direction
and the corresponding flux $\Phi_y$ is well-defined.
General non-contractible loops can be decomposed into a sum of $C_x$ or $C_y$
and contractible loops, and corresponding fluxes are clearly given.

\section{LSM theorem for site-dependent spin system}
\label{app:non-uniform}

The LSM theorems can be extended to systems with non-uniform spins $S_j$
depending on spatial positions.
Let us consider an XXZ model with an even number of half-integer spins and an arbitrary number
of integer spins.
(If the number of half-integer spins is odd, there is a Kramers degeneracy in presence of the time-reversal symmetry.)
For simplicity, we consider the square lattice $\Lambda$ with linear size $L$
which has the $C_{2v}=\{1,C_2,M_{1},M_{2}\}$ symmetry about a site with a half-integer spin 
(and the mirror planes contain that site).
Then, Claim \ref{thm2} holds for such a model as well,
because the phase factor in the commutation relations of the combined symmetry operators
is $\exp(i2\pi \sum_{x,y\neq0}S_j^z)=-1$ in this case.
Similarly, Claim \ref{thm2TR} also holds since $(\tilde{C}_2)^2=-1$.
We can repeat the same proof as in the main text, which is omitted here.

\begin{claim}
Consider an XXZ model with site-dependent spins and suppose that the model
has the on-site ${\mathrm U}(1)\rtimes{\mathbb Z}_2$ symmetry and point group $C_{2v}$ symmetry about a site
with a half-integer spin.
Then, the ground state cannot be uniquely gapped under the periodic boundary condition.
\end{claim}
\begin{claim}
Consider an XXZ model with site-dependent spins and suppose that the model
has the on-site ${\mathrm U}(1)\times{\mathbb Z}_2^T$ symmetry and $C_2$ rotation symmetry about a site
with a half-integer spin.
Then, the ground state cannot be uniquely gapped under the periodic boundary condition.
\end{claim}

\section{LSM theorem by local flux twisting}
\label{sec:local}
Point group symmetry can be twisted not only by a uniform U(1) flux but also by
a local flux.
In the latter case, it turns out that a small local flux is not sufficient to twist the symmetry
and a $\pi$-flux is required to obtain an LSM theorem.
Similarly to the uniform flux case,
we can show existence of a pair of the eigenstates for the twisted Hamiltonian as discussed previously~\cite{Po2017},
but it is difficult to extrapolate it to the untwisted Hamiltonian in a controlled way.
Just for a comparison, here we briefly discuss the LSM theorem by a local flux.
The statements are as follows.
\begin{claim}
A two-dimensional
half-integer spin
XXZ model with the on-site ${\mathrm U}(1)\rtimes{\mathbb Z}_2$ symmetry and 
the $C_2$ rotation symmetry
does not have a unique gapped ground state under the periodic boundary condition.
\label{conj1}
\end{claim}
\begin{claim}
A two-dimensional
half-integer spin nearest neighbor interacting
system with the on-site ${\mathbb Z}_2\times{\mathbb Z}_2^T$ symmetry and 
the $C_2$ rotation symmetry
does not have a unique gapped ground state under the periodic boundary condition.
\label{conj2}
\end{claim}

These claims look similar to the previous theorems, but there are some differences.
We require only one symmetry operation, the $C_2$ rotation symmetry,
for Claim~\ref{conj1} and ~\ref{conj2}.
In this sense, 
they are stronger than Claims~\ref{thm2}, 
althhough it is difficult to extrapolate these claims to the untwisted Hamiltonian.

\begin{figure}[tbh]
\includegraphics[width=8.0cm]{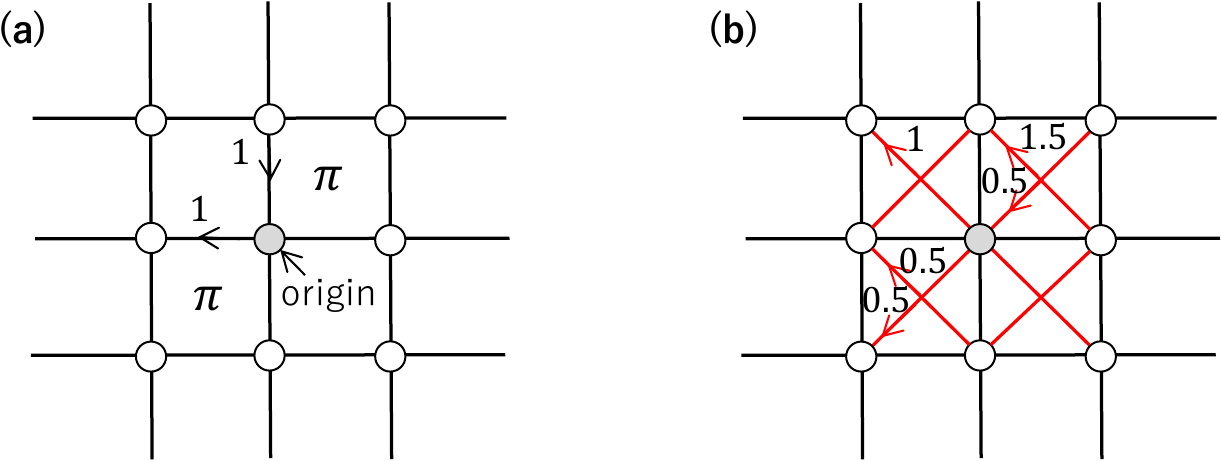}
\caption{The gauge field of the local $\pi$-fluxes on the square lattice for (a) the nearest neighbor interaction
and (b) the next nearest neighbor interaction.
The numbers are in unit of $\pi$.
The center site is regarded as the origin.
}
\label{fig:local}
\end{figure}

We briefly discuss ground state degeneracy of the twisted Hamiltonian in a way slightly
different from the previous study~\cite{Po2017}.
We consider a half-integer spin Hamiltonian Eq.~\eqref{eq:H2D0} on the two-dimensional square lattice 
with $C_2$-rotational symmetry about the origin for simplicity.
Extensions to other lattices are briefly touched on later.
To make the on-site symmetry and $C_2$ rotation symmetry correlated,
we introduce local $\pi$-fluxes only around the origin as shown in Fig.~\ref{fig:local}
and change $J_{jk}\to J_{jk}e^{iA_{jk}}$ correspondingly.
Note that ${\mathrm U}(1)\rtimes {\mathbb Z}_2$ symmetry rather than ${\mathbb Z}_2\times{\mathbb Z}_2$
is required so that the gauge field for the next nearest neighbor bonds (or further neighbor interactions) is well defined.
Then, it is easy to see that the $C_2$ symmetry gets twisted by the local flux as
\begin{align}
\tilde{C}_2&=C_2u, \quad u=e^{i\pi S_0^z},
\end{align}
where $S_0^z$ is the spin operator at the origin.
The projective operator $\tilde{C}_2$ has a non-trivial commutation relation with 
the $R^x_{\pi}$-operator,
\begin{align}
\tilde{C}_2R^x_{\pi}=-R^x_\pi\tilde{C}_2.
\end{align}
Therefore, every eigenvalue of $H(\phi=\pi)$ is degenerate when the system has the spin rotation symmetry
and the $C_2$ rotation symmetry.
We note that the mirrors $M_x,M_y,M_{xy},M_{\bar{xy}}$ do not lead to ground state degeneracy. 
For example, one can consider a simple dimerization Hamiltonian with $M_y$-symmetry, 
$H=\sum_{y_i}\sum_{x_i:{\rm even}} J\vec{S}_{i}\cdot\vec{S}_{i+\hat{x}}$,
which has a unique gapped ground state. 
Similarly, a dimerization Hamiltonian only with the next nearest neighbor interaction parallel to the $x=y$ line
has the $M_{xy}$ symmetry, but its ground state is unique.
For a general interaction $J_{jk}$, there is no twisted mirror symmetry operator
in presence of the local flux.

The above argument is valid also for the XYZ spin model when the interactions are only for the nearest
neighbor sites, because 
only the ${\mathbb Z}_2\times{\mathbb Z}_2$ spin-rotation symmetry is used in this case.
We can
add a nearest neighbor symmetric interaction to the XYZ spin model,
\begin{align}
H(\phi)&=H_{XYZ}(\phi)+H_{\Gamma}(\phi),\\
H_{XYZ}&=\sum_{jk} J^x_{jk}e^{iA_{jk}}S^x_jS^x_k+J^y_{jk}e^{iA_{jk}}S^y_jS^y_k+J^z_{jk}S^z_jS^z_k,\\ 
H_{\Gamma}&=\sum_{jk} \Gamma_{jk}e^{iA_{jk}}(S^x_jS^y_k+S^y_jS^x_k),
\end{align}
where $A_{ij}=0,\pi$ is the ${\mathbb Z}_2$ gauge field as  shown in Fig.~\ref{fig:local} (a).
We suppose that the system has $C_2$-symmetry in absence of a flux.
The Hamiltonians $H(\phi=0), H(\phi=\pi)$
have spin $\pi$-rotation symmetry about $z$-axis $R^z_{\pi}$ and
time-reversal symmetry $T=R^y_{\pi}K$.
In presence of $\phi=\pi$ (Fig.~\ref{fig:local}), 
the twisted $C_2$-operator $\tilde{C}_2=C_2u$
commutes with the time-reversal symmetry operator,
\begin{align}
T\tilde{C}_2=\tilde{C}_2T.
\end{align}
Note that $(\tilde{C}_2)^2=-1$ and its eigenvalues are $c=\pm i$, while $T^2=+1$ in our system with 
an even $L$.
Therefore,
similarly to the discussions in the previous sections, 
every eigenstate of $H(\phi=\pi)$ is degenerate thanks to 
the above commutation relation with the unitary operator $\tilde{C}_2$ and the anti-unitary
operator $T$.

\begin{figure}[tbh]
\includegraphics[width=8.0cm]{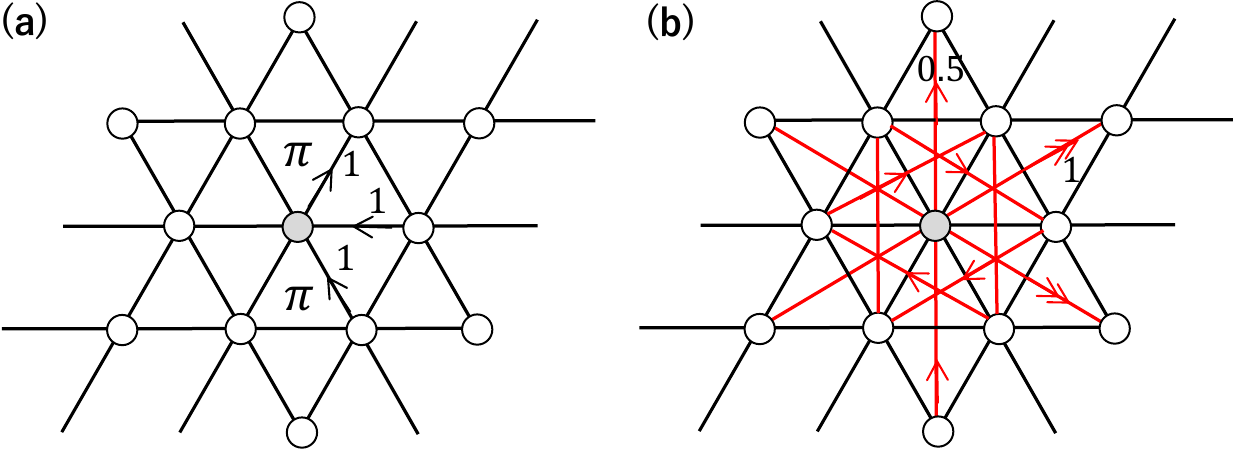}
\caption{The gauge field of the local $\pi$-fluxes on the triangular lattice for (a) the nearest neighbor interaction
and (b) the next nearest neighbor interaction.
The numbers are in unit of $\pi$.
}
\label{fig:localtri}
\end{figure}

We can extend our argument to general lattices with the $C_2$ rotation symmetry about a site.
For example, Fig.~\ref{fig:localtri} shows a gauge field configuration for the triangular lattice and clearly 
the same argument applies for this case.
On the other hand, for the honeycomb lattice,
we cannot realize non-trivial commutation relations of 
symmetry operators with the $C_3$ rotation symmetry about a site.
The twisted $C_3$ symmetry operator itself is well-defined as $\tilde{C}_3=C_3u_3$ with $u_3=e^{i2\pi/3S_0^z}$
in presence of $2\pi/3$-fluxes on three hexagons surrounding the origin.
but its commutation relation is not closed in $\{\tilde{C}_3,R_{\pi}^x\}$
and degeneray cannot be derived from the symmetry.

The next step for the LSM theorem is to extrapolate the twisted system to the untwisted one.
Unfortunately, in contrast to the tiny uniform flux in the main text, 
it is non-trivial whether or not the degeneracy can remain even for the 
untwisted Hamiltonian $H(\phi=0)$,
because a change in the Hamiltonian for a local $\pi$-flux is $O(1)$ 
with respect to the system size.
This problem is reduced to the spectral robustness against the U(1) twisting in a one dimensional system,
when the interaction in one of the $x$- or $y$-direction is turned off.
(There are $L-1$ untwisted one dimensional chains and a $\pi$-twisted chain.)
Even the one dimensional problem has not been fully resolved as mentioned previously,
and the present problem would be more complicated.
However, we naively expect that
an untwisted system without a local flux has a unique gapped ground state,
if and only if
the corresponding twisted system with a local flux has a unique gapped ground state.
A dangerous possibility is that the local flux creates a local gapless bound state,
but if exist, there should be two bound modes located at nearby plaquettes with the $\pi$-fluxes
and they will hybridize to get gapped.
If this assumption is correct, Claims~\ref{conj1} and \ref{conj2} will hold.
 
A similar hypothesis has been assumed to hold in U(1) symmetric systems, 
and numerical calculations suggest that the hypothesis is indeed correct~\cite{Hatsugai2006,Hirano2008}.
On the other hand, 
the above hypothesis might not hold if the plaquettes with the fluxes
are located far apart each other.
Such a situation has been discussed in the previous study~\cite{Po2017}.
In this case, there could be isolated bound modes around each flux where overlap between
the bound modes is vanishing.
Therefore, in general, the ground state degeneracies for $H(0)$ and $H(\phi\neq0)$ might be different,
when the local fluxes are introduced in an arbitrary distance.

\bibliography{ref}

\end{document}